\documentclass[journal]{IEEEtran}
\usepackage[cmex10]{amsmath}
\usepackage{amssymb}
\usepackage{epsfig}
\usepackage{subfigure}
\usepackage{cite}
\usepackage{graphicx}
\usepackage{color}
\usepackage{bm}
\usepackage{booktabs}
\usepackage{gensymb}
\usepackage{mathrsfs}
\usepackage{xfrac}
\usepackage{algorithm}
\usepackage{algorithmic}
\newlength{\figwidth}
\makeatletter
\begin{document}
%
\title{Energy-Delay Tradeoff in Helper-Assisted NOMA-MEC Systems: A Four-Sided Matching Algorithm}
%

%

\author{Mengmeng~Ren,
        ~Long~Yang,
        ~Hai~Jiang,~Jian~Chen,
        and~Yuchen~Zhou 
\thanks{This paper was partially presented at the IEEE WCNC 2021 \cite{My_WCNC}.} 
\thanks{M. Ren, L. Yang, J. Chen, and Y. Zhou are with the State Key Laboratory of Integrated Services Networks, Xidian University, Xi'an 710071, China (e-mail: renmengmeng@stu.xidian.edu.cn;  lyang@xidian.edu.cn; jianchen@mail.xidian.edu.cn; ychenzhou@163.com).}
\thanks{H. Jiang is with the Department of Electrical and Computer Engineering, University of Alberta, Edmonton, AB T6G 1H9, Canada (e-mail: hai1@ualberta.ca).}
\vspace{-2em}
}

%
%

\markboth{Journal of \LaTeX\ Class Files,~Vol.~14, No.~8, August~2015}%
{Shell \MakeLowercase{\textit{et al.}}: Bare Demo of IEEEtran.cls for IEEE Journals}
%



\maketitle


\begin{abstract}
This paper designs a helper-assisted resource allocation strategy in non-orthogonal multiple access (NOMA)-enabled mobile edge computing (MEC) systems, in order to guarantee the quality of service (QoS) of the energy/delay-sensitive user equipments (UEs).
To achieve a tradeoff between the energy consumption and the delay, we introduce a novel performance metric, called \emph{energy-delay tradeoff}, which is defined as the weighted sum of energy consumption and delay.
The joint optimization of user association, resource block (RB) assignment, power allocation, task assignment, and computation resource allocation is formulated as a mixed-integer nonlinear programming problem with the aim of minimizing the maximal energy-delay tradeoff.
Due to the non-convexity of the formulated problem with coupled and 0-1 variables, this problem cannot be directly solved with polynomial complexity.
To tackle this challenge,
we first decouple the formulated problem into a power allocation, task assignment and computation resource allocation (PATACRA) subproblem. Then, with the solution obtained from the PATACRA subproblem, we equivalently reformulate the original problem as a discrete user association and RB assignment (DUARA) problem. For the PATACRA subproblem, an iterative parametric convex approximation (IPCA) algorithm is proposed. Then, based on the solution obtained from the PATACRA subproblem,
we first model the DUARA problem as a four-sided matching problem, and then propose a low-complexity four-sided UE-RB-helper-server matching (FS-URHSM) algorithm.
Theoretical analysis demonstrates that the proposed algorithms are guaranteed to converge to stable solutions with polynomial complexity. Finally, simulation results are provided to show the superior performance of our proposed algorithm in terms of the energy consumption and the delay.
\end{abstract}


%
\IEEEpeerreviewmaketitle

\section{Introduction}
%
%
%
%
\IEEEPARstart{D}{riven} by the proliferation of fifth generation (5G) and beyond networks, a variety of intelligent applications and services have emerged, coexisting with a large number of delay-sensitive and computation-sensitive tasks\cite{IoTSerZ,IoTMEC,2019Tut2}. As a result, users may face great challenges in computation/processing capability due to their limited computation/storage resources and battery capacity.
Conventionally, mobile cloud computing (MCC) is a promising technology to serve resource-constrained users by using abundant computation resources deployed at central cloud servers \cite{Tut1}. However, as the central cloud is generally distant from users, MCC may incur excessive delay due to the long-distance transmission over backhaul/core-networks. To tackle this issue, mobile edge computing (MEC) has been proposed to provide cloud-like services by deploying servers at the network edge, called MEC servers, with which multiple resource-constrained users will be able to wirelessly offload their tasks to MEC servers in close proximity \cite{IoTMEC2,TDMA2,2019Tut3}. Therefore, assisted by proximate servers, resource-constrained users can offload and complete tasks with much lower delay than the counterpart in MCC\cite{IoTMEC1}.  

In MEC systems, according to the separability of tasks, the task offloading can be classified into two categories: \emph{inseparable task offloading} \cite{NOMAMEC_DZG,SYP_TWC_OMA,ZFU_OMA_MEC} and \emph{separable task offloading} \cite{OMA_SingleServer_TCOM,OMA_MultiServer_TGCN}. In the \emph{inseparable task offloading}, each user attempts to offload the entire task to the MEC server. Specifically, recent works investigated on the communication and computation resources allocation to minimize the task completion delay \cite{NOMAMEC_DZG,SYP_TWC_OMA} or maximize the computation rate \cite{ZFU_OMA_MEC}. In \cite{NOMAMEC_DZG}, two algorithms based on Dinkelbach's and Newton's methods were proposed to minimize the overall delay for two-user systems supported by a MEC server. Further, for multi-server MEC systems, the overall delay of all users was minimized by jointly optimizing the user association and resource allocation \cite{SYP_TWC_OMA}. Moreover, under the delay constraint, the weighted sum computation rate of all users was maximized by a proposed iterative algorithm \cite{ZFU_OMA_MEC}. In the \emph{separable task offloading}, each user offloads several portions of the task to the servers while locally executing the remaining portion, in order to fully utilize the local computation resource of users and relief the burden of task offloading. For users supported by single server, 
an enhanced online Lyapunov optimization algorithm was proposed to maximize the long-term average data rate\cite{OMA_SingleServer_TCOM}. Further, for users supported by multiple servers, the long-term energy consumption of all users was minimized by an enhanced deep reinforcement learning approach \cite{OMA_MultiServer_TGCN}.

All aforementioned works \cite{NOMAMEC_DZG,SYP_TWC_OMA,ZFU_OMA_MEC,OMA_SingleServer_TCOM,OMA_MultiServer_TGCN} focused on enhancing the performance of task offloading by properly utilizing the inherent computation/communication resources offered by the users/MEC servers. In fact, due to the burst nature of users' traffics, each user is very likely to be surrounded by some idle-state users with spare computation/communication resources, especially in practical 5G and its beyond systems that support massive users. Thus, if recruiting those idle-state users as \emph{helpers} to assist the task offloading/execution, the performance of task offloading can be further improved by utilizing the spare resources offered by idle-state users.
Inspired by this fact, the helper-assisted MEC has been investigated with using the orthogonal multiple access (OMA)-enabled offloading, i.e., \emph{helper-assisted OMA-MEC} \cite{TDMAD2D6,FDMAD2D1,helper_TVT_FDMA,helper_FDMA1}. In \cite{TDMAD2D6}, a heuristic task assignment scheme was proposed to minimize the overall delay, with which a typical user sequentially offloads tasks to multiple helpers via time-division multiple access (TDMA). On the other hand, by employing frequency-division multiple access (FDMA) for task offloading, the number of served users was maximized in \cite{FDMAD2D1}, where each user occupies two subchannels for offloading to the MEC server and the helper, respectively.
Further, with the integration of TDMA and FDMA, the work in \cite{helper_TVT_FDMA} and \cite{helper_FDMA1} investigated hybrid TDMA/FDMA offloading schemes to minimize the energy consumption. Specifically, a helper-assisted offloading scheme was proposed in \cite{helper_TVT_FDMA}, with which users offload tasks to the MEC server assisted by a selected helper.
Likewise, a two-level alternative method was proposed to jointly optimize the offloading decision, computation and communication resources allocation \cite{helper_FDMA1}. 

{In practice, since helpers may be geographically closer to users than MEC servers, they can easily form a near-far effect with MEC servers, thus naturally providing possibility for implementing the non-orthogonal multiple access (NOMA)-enabled task offloading. Inspired by this, the NOMA technique can be integrated into the helper-assisted MEC, termed as \emph{helper-assisted NOMA-MEC}. Specifically, each user can simultaneously offload subtasks to the helpers and the MEC servers by using power-domain multiplexing. Compared with the helper-assisted OMA-MEC, the helper-assisted NOMA-MEC can achieve higher spectral efficiency and lower offloading/computing delay \cite{TDMAD2D6,FDMAD2D1,helper_TVT_FDMA,helper_FDMA1}. More recently, the helper-assisted NOMA-MEC has been investigated with using a single helper \cite{3nodes_NOMA,3nodes_NOMA_TVT} or multiple helpers \cite{MultipleHelper_NOMAMEC,My_helpers}. For the single-user NOMA-MEC assisted by only one helper, the minimization of energy consumption was investigated in \cite{3nodes_NOMA} and \cite{3nodes_NOMA_TVT}. In \cite{3nodes_NOMA}, an alternative iterative algorithm was proposed to jointly optimize the delay, task splitting, and power allocation.
As an extension of \cite{3nodes_NOMA}, the computation frequencies of the user and the helper were further jointly optimized in \cite{3nodes_NOMA_TVT}. When multiple helpers are available, an interior-point-based algorithm was proposed in \cite{MultipleHelper_NOMAMEC} to maximize the offloading data of one task-driven user. In \cite{My_helpers}, an adaptive computing strategy with a helper scheduling scheme was proposed, in which one helper is adaptively scheduled from multiple helpers to assist task offloading.}

From above works \cite{3nodes_NOMA,3nodes_NOMA_TVT,MultipleHelper_NOMAMEC,My_helpers} on the helper-assisted NOMA-MEC, it can be observed that the helper-assisted task offloading was investigated for only a typical user. However, in multiuser NOMA-MEC systems, it is very likely that some active users require computing services at the same time, indicating that their tasks need to be offloaded/executed simultaneously. In this situation, \textit{how to fairly serve them is still an open issue for the helper-assisted NOMA-MEC.} Moreover, in practical multiuser MEC systems driven by LTE/5G, multiple orthogonal time-frequency resource blocks (RBs) are assigned to different users, indicating that an appropriate RB allocation can further improve the performance of offloading and computation. Note that existing works \cite{TDMAD2D6,FDMAD2D1,helper_TVT_FDMA} on helper-assisted OMA-MEC mainly investigated the association between users and servers. \textit{Therefore, how to jointly optimize the allocation of RBs and the association among users, helpers, and servers to improve the computation performance is also uninvestigated.}


Motivated by the above observations, this paper develops a novel paradigm for multi-helper assisted NOMA offloading.
To achieve a tradeoff between the energy consumption and the delay, we introduce a novel performance metric, called \emph{energy-delay tradeoff (EDT)}, which is defined as the weighted sum of the energy consumption and the delay. Then, from the perspective of min-max fairness, we formulate the joint optimization of user association, RB assignment, power allocation, task assignment, and computation resource allocation to minimize the maximal EDT (mEDT) among user equipments (UEs). Since the formulated problem is a mixed-integer nonlinear programming (MINLP) problem with coupled discrete and continuous variables, it is challenging to design an algorithm with low complexity and low performance loss. The main contributions of this paper are summarized as follows.

\begin{itemize}
\item \textit{Equivalent problem transformation scheme:} We design an equivalent transformation scheme to solve the formulated MINLP problem. Specifically, under the condition of given user association and RB assignment (UARA), the original problem will be reduced to a subproblem with only continuous variables (i.e., power allocation, task assignment, and computation resource allocation), called PATACRA subproblem. Then, regarding the continuous variables as functions of the UARA, we equivalently reformulate the original problem as a discrete UARA (DUARA) problem.
\item \textit{Alternating-free iteration algorithm for PATACRA subproblem:} For the non-convex PATACRA subproblem, we introduce auxiliary variables to transform it into a form that can be approximated appropriately, and propose an iterative parametric convex approximation (IPCA) algorithm. Unlike existing alternating-based algorithms \cite{ZZ_CL_AO,ZZ_CL} that require further decomposition of the non-convex problem, the proposed IPCA algorithm can be directly applied to the non-convex PATACRA subproblem with low complexity.
\item \textit{Novel four-sided matching algorithm for DUARA problem:} For the non-convex DUARA problem, we first model it as a four-sided matching problem with the aim of finding the best matching with the highest objective value. To solve this matching problem, we propose a novel four-sided UE-RB-helper-server matching (FS-URHSM) algorithm, where a sequential swap operation and a leaving and joining-in operation are designed. Unlike conventional deferred-acceptance/swap-operation based algorithms\cite{3D_TMC_cyclic,huang2010circular,3DMatching_Zhang} that are only feasible for two/three-sided matching problems, our proposed FS-URHSM algorithm can efficiently generate a feasible and stable four-sided matching.
\item \textit{ Convergence and stability analysis: } To demonstrate the convergence of the proposed FS-URHSM algorithm, we first prove that the proposed IPCA algorithm will converge to a stationary point. Then, based on the convergence of the IPCA algorithm, we theoretically show that the FS-URHSM algorithm is guaranteed to converge to a stable matching with polynomial complexity, which is much less than the exponential complexity required by exhaustively searching all possible matching conditions.
\end{itemize}

The rest of this paper is organized as follows. Section \ref{Section_system} describes the helper-assisted NOMA-MEC system, along with the helper-assisted offloading model and the task computing model.
Section \ref{Section_formulation} formulates the joint optimization problem to minimize the mEDT, and equivalently reformulates the original problem as a DUARA problem. Section \ref{SCA_section} proposes the IPCA algorithm to solve the PATACRA subproblem and analyzes the convergence and complexity. To solve the DUARA problem, Section \ref{3DMatching} models this problem as a four-sided matching problem and proposes a FS-URHSM algorithm. Then, the convergence and complexity are analyzed. Section \ref{section_simulation} illustrates the numerical results, followed by Section \ref{Section_conclusion} concludes this paper.
\section{System Model}
\label{Section_system}
In this section, we first introduce the helper-assisted NOMA-MEC system, and then describe the helper-assisted offloading model and the task computing model.
\subsection{Helper-Assisted NOMA-MEC System}
\begin{figure}[!t]
\centering{\includegraphics[scale=0.44]{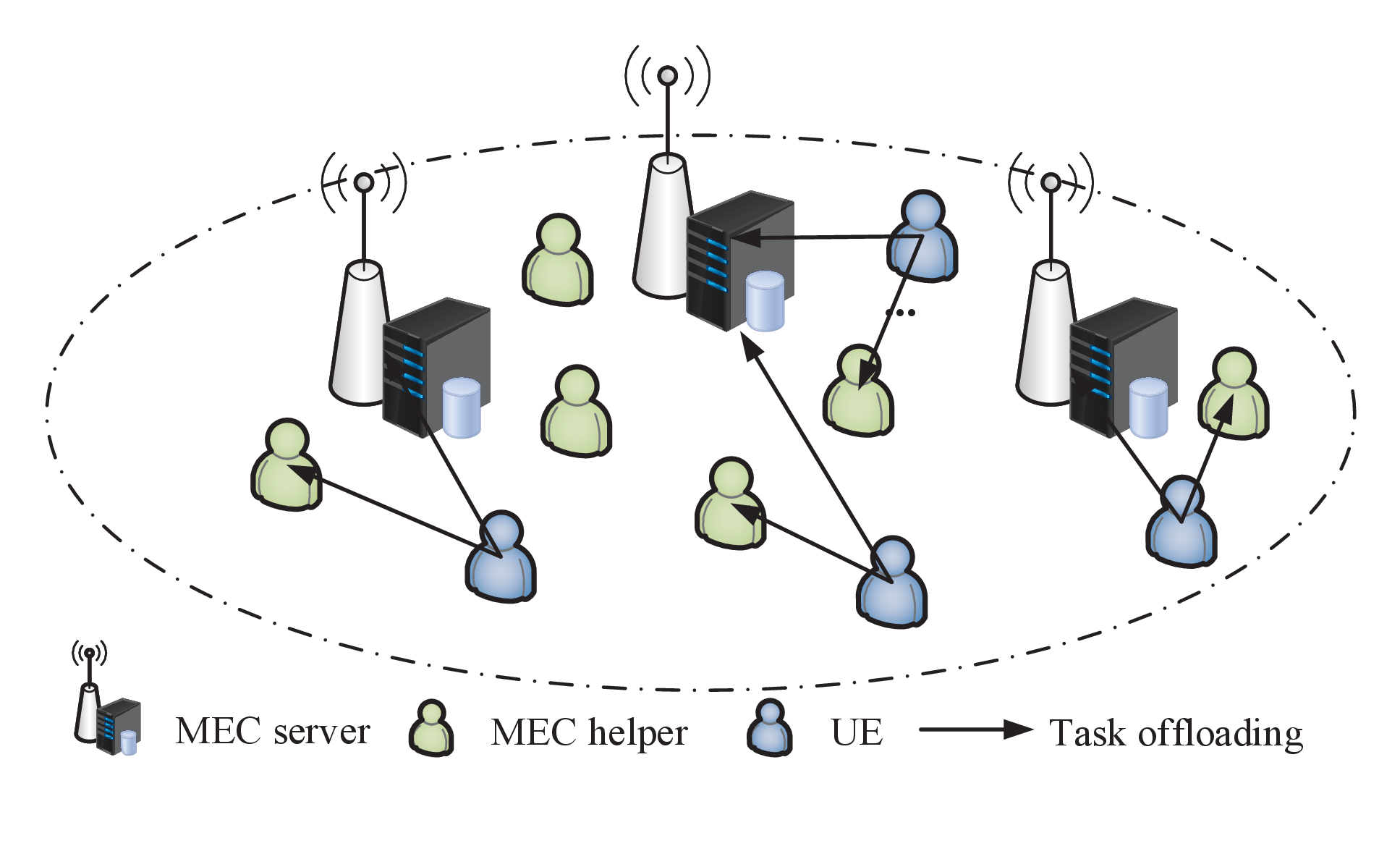}} 
\caption{Helper-assisted NOMA-MEC system.}
\vspace{-1em}
\label{system1}
\end{figure}

As shown in Fig. \ref{system1}, we consider a helper-assisted NOMA-MEC system consisting of $N$ task-driven UEs, $M$ active MEC helpers, and $K$ MEC servers. Let $U_n$ ($n \in \mathcal{N} \triangleq \{1,\ldots,N\}$), $H_m$ ($m \in \mathcal{M}\triangleq \{ 1,\ldots,M \}$) and $S_k$ ($k \in \mathcal{K}\triangleq \{ 1,\ldots,K \}$) denote the sets of UEs, helpers, and servers, respectively.
We assume that each UE, say $U_n$, has a task to be executed, represented by $\bm{L}_n \triangleq \{D_n, I_n, T _n^{\max}, \omega_{n}^E, \omega_{n}^T \}$, where $D_n$ denotes the input data size of the task, $I_n$ represents the computation intensity of the task (i.e., the required CPU cycles for computing $1$-bit data), $T _n^{\max}$ denotes the maximum delay tolerance, $\omega_{n}^E$ and $\omega_{n}^T$ represent the weight factors to measure the energy consumption and delay sensitivity with satisfying $\omega_{n}^E+\omega_{n}^T=1$, respectively.
Since the decoding complexity of the successive interference cancellation (SIC) technique in practical downlink NOMA systems will exponentially increase with the number of the receivers (i.e., helpers/servers) \cite{Multi_NOMAMEC_SM,FF_TCOM_imCSI}, we assume that each UE is associated with one server and one helper to offload subtasks by employing downlink NOMA. Let $x_{n,k}$ denote the association indicator among UEs and servers, where $x_{n,k} = 1 $ represents that server $S_k$ is associated with UE $U_n$; otherwise, $x_{n,k} = 0$.
We assume that each server serves at most $N_k^{\max}$ UEs, which is expressed as
\begin{align}
\label{US_A1}
\sum\limits_{n=1}^N { {x_{_{n,k}} \leq N_k^{\max}} } ,\forall k \in \mathcal{K}.
\end{align}
Each UE is served by one server, which is expressed as
\begin{align}
\label{US_A2}
\sum\limits_{k = 1}^K { {x_{_{n,k}} = 1} } ,\forall n \in \mathcal{N}.
\end{align}
Let $y_{n,m}$ denote the association indicator among UEs and helpers, where $y_{n,m} = 1 $ represents that helper $H_m$ is associated with UE $U_n$; otherwise, $y_{n,m} = 0$.
We assume that each helper only serves at most one UE, expressed as
\begin{align}
\label{Helper_UE_e1}
\sum\limits_{n = 1}^N { {y_{_{n,m}} \leq 1} } ,\forall m \in {\mathcal{M}}.
\end{align}
Each UE is served by one MEC helper, and the association variables are constrained by
\begin{align}
\label{Helper_UE_e2}
\sum \limits_{m = 1}^M y_{_{n,m}} = 1 ,\forall n \in {\mathcal{N}}.
\end{align}
Moreover, to avoid the inter-UE interference, we assume that UEs occupy the orthogonal RBs for task offloading. More specifically, the available spectrum is divided into $L$ ($L\geq N$) orthogonal RBs, denoted by $RB_l$, $ l \in \mathcal{L} \triangleq \{1,\ldots, L \}$. Let $z_{n,l}$ denote the RB assignment indicator, where $z_{n,l} = 1$ denotes that UE $U_n$ is matched to RB $RB_l$; otherwise, $z_{n,l} = 0$. Practically, each UE can be allocated to one RB, which is expressed as
\begin{align}
\label{RB_UE_e1}
\sum\limits_{l = 1}^L {{z_{n,l}} = 1,} \forall n \in {\mathcal{N}}.
\end{align}
Each RB is allocated to at most one UE, which is constrained by
\begin{align}
\label{RB_UE_e2}
\sum\limits_{n = 1}^N {{z_{n,l}} \leq 1,} \forall k \in {\mathcal{L}}.
\end{align}

With the above constraints of user association \eqref{US_A1}$\sim$\eqref{Helper_UE_e2} and RB allocation \eqref{RB_UE_e1} and \eqref{RB_UE_e2}, for each UE (called $U_n$), it will be assigned to a helper, a server, and an RB, which can be denoted as $H_m$, $S_k$, and $RB_l$, respectively. Before task offloading, UE $U_n$ will divide task $\bm{L}_n$ into three subtasks: 1) subtask $\varphi_{n}^{H}$ with $\eta_{n}^{H} D_n$ bits; 2) subtask $\varphi_{n}^{S}$ with $\eta_{n}^{S} D_n$ bits; 3) subtask $\varphi_{n}^{L}$ with $(1-\eta_{n}^{H} -\eta_{n}^S)D_n$ bits, where $\eta_{n}^{H}$ and $\eta_{n}^{S} $ denote the task splitting coefficients for helper $H_{m}$ and server $S_{k}$, respectively. The task splitting coefficient satisfies $ 0 \leq \eta _{n}^H, \eta_n^S \leq 1$ and $ \eta _{n}^H+\eta_n^S \leq 1$ ($\forall n \in \mathcal{N}$). After the task is divided, UE $U_n$ will offload subtasks $\varphi_{n}^{H}$ and $\varphi_{n}^{S}$ to associated helper $H_m$ and server $S_k$ by employing the downlink NOMA, while executing subtask $\varphi_{n}^{L}$ locally during this offloading phase. Then, after successfully decoding the desired subtasks, MEC helper $H_m$ and server $S_k$ will execute the decoded subtasks. Since the computing results only include a few bits, it is reasonable to assume that the duration of the result downloading is negligible\cite{NOMAMEC_FF_TCOM,FF_TCOM_imCSI}. Therefore, we generally pay more attention to the design of optimization strategies from the perspective of users. Note that the details of the helper-assisted offloading model and the task computing model are described as follows.
\subsection{Helper-Assisted Offloading Model}
In the considered system, all the UEs/helpers/servers are equipped with a single antenna and operate in half duplex mode. Without loss of generality, we assume that all the wireless channels experience independent but non-identically Rayleigh block-fading distribution, which means that the channel gains remain unchanged in one transmission block but may vary independently over different blocks\cite{NOMA_YL}. The channel gains of links from UE $U_n$ to MEC helper $H_m$ and server $S_k$ over RB $RB_l$ can be denoted as $|h_{n, m,l}^H|^2 $ and $|h_{n, k,l}^S|^2 $, respectively. Following the same assumption in \cite{MultipleHelper_NOMAMEC,FF_TCOM_imCSI,3nodes_NOMA,3nodes_NOMA_TVT}, we assume that SIC can be perfectly performed when implementing the downlink NOMA. Considering the relationship between channel gains $|h_{n,m,l}^H|^2$ and $|h_{n,k,l}^S|^2$, the decoding order at the helper and the MEC server can be classified into the following two cases: 1) If $|h_{n,m,l }^H|^2 \geq |h_{n,k,l}^S|^2$, helper $H_m$ can use the SIC technology to eliminate subtask $\varphi_n^S$ that sent to the server, and then decode its desired subtask $\varphi_n^H$. At this point, server $S_k$ will decode subtask $\varphi_n^S$ with regarding $\varphi_n^H$ as the interference. 2) If $|h_{n,m,l }^H|^2 < |h_{n,k,l}^S|^2$, server $S_k$ will decode $\varphi_n^S$ after subtracting $\varphi_n^H$, while helper $H_m$ will decode $\varphi_n^H$ with regarding $\varphi_n^S$ as the interference. Therefore, helper $H_m$ will decode subtask $\varphi_n^H$ with the following data rate
\begin{equation}
\label{e3_1}
r_{n,m,l}^H= B \log _2 (1+ \frac{p_{n}^H g_{n,m,l}^H}{\text{O}_{n,k,m,l} p_{n}^S g_{n,m,l}^H + 1} ),
\end{equation}
where $g^H_{n,m,l} \triangleq {|h^H_{n,m,l}| ^2}/{\sigma ^2}$, $\sigma ^2$ is the additive white Gaussian noise (AWGN) power, $B$ is the bandwidth of RB $RB_l$, $p_{n}^H$ denotes the transmit power of UE $U_n$ for helper $H_m$, and $\text{O}_{n,k,m,l} = 0$ if $|h_{n,m,l }^H|^2 \geq |h_{n,k,l}^S|^2$, otherwise $\text{O}_{n,k,m,l}= 1$. According to this data rate, the transmission delay of offloading subtask $\varphi_n^H$ is given by
\begin{equation}
\label{trans_delay_helper}
\tau^{\text{tran-H}}_{n} \triangleq \frac{\eta_{n}^H D_n }{\sum\limits_{m=1}^{M}\sum\limits_{l=1}^{L}y_{n,m}z_{n,l}r_{n,m,l}^H}.
\end{equation}
Likewise, the data rate for server $S_k$ to decode subtask $\varphi_n^S$ is given by
\begin{equation}
\label{e4_1}
r_{n,k,l} ^S=  B \log _2 \big(1+ \frac{p_{n}^S g_{n,k,l}^S } {(1-\text{O}_{n,k,m,l})p_{n}^H g_{n,k,l}^S +1 }\big),
\end{equation}
where $g^S_{n,k,l} \triangleq {|h^S_{n,k,l}| ^2}/{\sigma ^2}$, and $p_{n}^S$ is the transmit power of UE $U_n$ for server $S_k$.
Hence, the transmission delay for server $S_k$ to receive subtask $\varphi_{n}^S$ is given by
\begin{equation}
\label{trans_delay_AP}
\tau^{\text{tran-S}}_{n} \triangleq \frac{\eta_{n}^S D_n }{\sum\limits_{k=1}^{K}\sum\limits_{l=1}^{L}x_{n,k}z_{n,l}r_{n,k,l}^S}.
\end{equation}
To efficiently utilize the spectrum resources, pure NOMA is emplyed for task offloading. Therefore, the transmission delays from UE $U_n$ to helper $H_{m}$ and MEC server $S_k$ satisfy
\begin{align}
\label{delayeq}
\tau^{\text{tran-H}}_{n}=\tau^{\text{tran-S}}_{n}, \forall n \in \mathcal{N},
\end{align}
and the transmission energy consumed by UE $U_n$ is given by
\begin{equation}
\label{e7}
e^{\text{tran}}_{n} =   p_{n}^H \tau ^{\text{tran-S}}_{n}+  p_{n}^S \tau^{\text{tran-H}}_{n}.
\end{equation}

\subsection{Task Computing Model}
\subsubsection{Computing at Helpers}
After successfully decoding subtask $\varphi_n^H$, helper $H_m$ will execute it with the following computing delay
\begin{equation}
\label{Exe_delay_at_helper}
\tau^\text{exe-H}_{n} \triangleq \frac{\eta_{n}^H D_n I_n}{\sum_{m=1}^M y_{n,m}f_m^H},
\end{equation}
where $f_m^H$ is the CPU frequency of helper $H_m$. Since the helpers are generally low-power compared to the MEC servers, we consider the energy consumption required for computing subtasks on different helpers. Note that the energy consumption for helper $H_m$ to compute subtask $\varphi_n^H$ is given by
\begin{align}
e_{n,m}^\text{exe-H} = \eta_{n}^H D_n I_n \varsigma (f^H_m)^2,
\end{align}
where $\varsigma $ denotes the effective capacitance coefficient depending on the chip architecture \cite{NOMAMEC_FF_TCOM}.
\subsubsection{Computing at Servers}
After successfully decoding subtask $\varphi_n^S$, server $S_k$ will compute it with the following delay
\begin{equation}
\label{delay_AP_exe}
\tau_{n}^{\text{exe-S}} \triangleq \frac{\eta_{n}^S D_n I_n}{ \sum_{k=1}^K x_{n,k}f_{n,k}^S},
\end{equation}
where ${f_{n,k}^S}$ is the CPU frequency allocated to UE $U_n$. Since MEC servers usually have a continuous power supply, we do not consider the energy consumption of the server-side task execution, which is also a general assumption in existing studies \cite{3nodes_NOMA,3nodes_NOMA_TVT,MultipleHelper_NOMAMEC}.
\subsubsection{Computing at UEs}
UE $U_n$ ($n\in \mathcal{N}$) will simultaneously compute subtask $\varphi_{n}^L$ while offloading subtasks $\varphi_n^H$ and $\varphi_n^S$, and the local computing delay is given by
\begin{align}
\label{e11}
t^L_n = \frac{(1-  \eta_{n }^S -  \eta_{n }^H) D_n I_n}{f^L_n},
\end{align}
where $f^L_n$ is the computational CPU frequency of UE $U_n$. Moreover, the energy consumption of UE $U_n$ for local computing can be expressed as
\begin{align}
e^L_n = (1-   \eta_{n }^S - \eta_{n }^H) D_n I_n \varsigma (f^L_n)^2.
\end{align}

\section{Problem formulation and reformulation}
\label{Section_formulation}
To minimize the mEDT, this section first gives the definition of the mEDT, and then formulates the joint optimization problem of the user association, RB assignment, power allocation, task assignment, and computation resource allocation. Finally, the optimization problem can be equivalently reformulated as a DUARA problem with only discrete 0-1 variables.
\subsection{Maximal Energy-Delay Tradeoff}
Since each MEC helper/server executes the desired subtask after successfully decoding it, the delay of subtask completion at helper $H_m$ and server $S_k$ can be defined as
\begin{align}
\label{delay_Hm}
t^{\text{off-H}}_n \triangleq   \tau^{\text{tran-H}}_{n} +  \tau_{n}^\text{exe-H},
\end{align}
and
\begin{align}
\label{delay_Ak}
t^{\text{off-S}}_n \triangleq \tau^{\text{tran-S}}_{n} +  \tau_n^\text{exe-S},
\end{align}
respectively. Therefore, the task completion delay of $U_n$ can be defined as
\begin{align}
\label{delay_all}
T_n \triangleq \max \{ t_n^L, t^{\text{off-H}}_n, t^{\text{off-S}}_n \}.
\end{align}

Moreover, the corresponding energy consumption for UE $U_n$ to complete task $\bm{L}_n$ is defined as the sum of the local computing energy consumption and the transmission energy consumption, which is given by
\begin{align}
E_n \triangleq   e^{\text{tran}}_{n}+e^L_n.
\end{align}
Hence, the mEDT is defined as the maximal weighted sum of energy consumption and delay among all the UEs, which is given by
\begin{align}
\text{EDT}^{\text{max}} (\bm{X},\bm{Y},\bm{Z}, \bm{p}, \bm{\eta},\bm{f}) &\triangleq \max \limits_{n\in\mathcal{N}} \{\omega_n^E  E_n  + \omega_n^T T_n\},
\end{align}
where $\bm{X} = \{ x_{n,k} | \forall n \in \mathcal{N}, \forall k \in \mathcal{M} \}$ and $\bm{Y} = \{ y_{n,m} | \forall n \in \mathcal{N}, \forall m \in \mathcal{K} \}$ denote the user association variables, $\bm{Z} = \{ z_{n,l} | \forall n \in \mathcal{N}, \forall l \in \mathcal{L} \}$ is the RB assignment, $\bm{p} = \{p_{1}^H,\ldots, p_{N}^H;p_{1}^S,\ldots, p_{N}^S \} \in \mathbb{R}_+$ denotes the power allocation, $\bm{\eta} =\{\eta _{1}^H,\ldots, \eta _{N}^H;\eta_{1}^S, \ldots, \eta _{N}^S \} \in \mathbb{R}_+$ denotes the task assignment, and $\bm{f}=\{f_{n,k}^S | \forall n \in \mathcal{N}, \forall k \in \mathcal{K}\}$ denotes the computation resource allocation of the MEC servers.
\subsection{Problem Formulation}
From the perspective of min-max fairness, the joint optimization problem can be formulated to minimize the mEDT, which is given by
\begin{subequations}
\label{formulationProblem}
\begin{align}
\label{eq17} &\min\limits_{\bm{X},\bm{Y},\bm{Z}, \bm{p}, \bm{\eta},\bm{f} } \text{EDT}^{\text{max}}(\bm{X},\bm{Y},\bm{Z}, \bm{p}, \bm{\eta},\bm{f} ) \\
\label{11all}            &\quad\quad \text{s.t.} \quad  \eqref{US_A1}, \eqref{US_A2}, \eqref{Helper_UE_e1},\eqref{Helper_UE_e2},\eqref{RB_UE_e1}, \eqref{RB_UE_e2},  \eqref{delayeq}, \\
\label{energyofHelper}& \quad \quad \quad \quad  \sum\limits_{n=1}^N y_{n,m} e_{n,m}^{\text{exe-H}} \leq E_m^{\max}, \forall m \in {\mathcal{M}},\\
\label{11a}& \quad\quad\quad\quad   \eta_{n}^S +\eta_{n}^H \leq 1, \: \forall n \in \mathcal{N}, \\
\label{11b}& \quad\quad\quad\quad 0\leq \eta_{n}^S, \eta_{n}^H, \: \forall n \in \mathcal{N}, \\
\label{f_all_constraint} & \quad\quad\quad\quad \sum\limits_{n=1}^{N} x_{n,k}f_{n,k}^S \leq F_{k}^{\max}, \forall k \in \mathcal{K}, \\
\label{A_constraints}& \quad\quad\quad\quad x_{n,k} \in \left\{ {0,1} \right\}, \: \forall n \in \mathcal{N},\forall k \in \mathcal{K},\\
\label{11d}& \quad\quad\quad\quad y_{n,m} \in \left\{ {0,1} \right\}, \: \forall n \in \mathcal{N},\forall m \in \mathcal{M},\\
\label{11c}&\quad\quad\quad\quad  {z_{n,l}} \in \left\{ {0,1} \right\}, \: \forall n \in \mathcal{N},\forall l \in \mathcal{L},\\
\label{constrain_Tn}&\quad\quad\quad\quad T_n \le T_n^{\max}, \: \forall n \in \mathcal{N}, \\
\label{11m}&\quad\quad\quad\quad  0 \le p_{n }^H, p_{n }^S, \: \forall n \in \mathcal{N}, \\
\label{11n}&\quad\quad\quad\quad  p_n^H + p_n^S \leq P_n^{\max}, \forall n \in \mathcal{N}.
\end{align}
\end{subequations}
Constraints \eqref{energyofHelper}$\sim$\eqref{11n} are explained as follows. \eqref{energyofHelper} represents that the energy consumption at each helper should not exceed the maximum tolerance $E_m^{\max}$. \eqref{11a} and \eqref{11b} represent the range of task splitting coefficients of UEs. \eqref{f_all_constraint} means that the computation resources allocated by each sever to the served users should not exceed its maximum thresholds. Constraints \eqref{A_constraints}, \eqref{11d}, and \eqref{11c} are 0-1 constraints of the user association and RB assignment.
Constraint \eqref{constrain_Tn} indicates that the overall delay of executing the whole task of UE $U_n$ cannot exceed the maximum tolerance $T_n^\text{max}$.
\eqref{11m} and \eqref{11n} are the transmit power constraints.

\subsection{Equivalent Problem Reformulation}
The formulated optimization problem (i.e., problem \eqref{formulationProblem}) is a mixed-integer nonlinear programming problem due to the non-convex objective function and non-convex constraints \eqref{delayeq}, \eqref{A_constraints}, \eqref{11d}, \eqref{11c}, and \eqref{constrain_Tn}. To tackle this challenging problem \eqref{formulationProblem}, we equivalently reformulate it as follows.

Given the user association $\bm{X}$ and $\bm{Y}$, and RB assignment $\bm{Z}$, problem \eqref{formulationProblem} can be decoupled as a PATACRA subproblem with respect to only continuous variables, which can be expressed by
\begin{subequations}
\label{CCRA}
\begin{align}
\label{eqCCRA} &\min\limits_{\bm{p}, \bm{\eta}, \bm{f}} \text{EDT}^{\text{max}}(\bm{p}, \bm{\eta}, \bm{f}) \\
\label{CCRO1}        &\quad\quad \text{s.t.} \quad  \eqref{delayeq}, \eqref{energyofHelper} \sim \eqref{f_all_constraint}, \eqref{constrain_Tn}\sim\eqref{11n}.
\end{align}
\end{subequations}
As user association $\bm{X}, \bm{Y}$ and RB assignment $\bm{Z}$ are assumed to be known in PATACRA subproblem, the solution obtained form PATACRA subproblem can be regarded as functions of the 0-1 variables $\{ \bm{X}, \bm{Y},\bm{Z} \}$ denoted as $\{ \bm{p}^*(\bm{X},\bm{Y},\bm{Z}),  \bm{\eta}^*(\bm{X},\bm{Y},\bm{Z}), \bm{f}^*(\bm{X},\bm{Y},\bm{Z})\}$. Therefore, by applying $\{ \bm{p}^*(\bm{X},\bm{Y},\bm{Z}),  \bm{\eta}^*(\bm{X},\bm{Y},\bm{Z}), \bm{f}^*(\bm{X},\bm{Y},\bm{Z})\}$ into problem \eqref{formulationProblem}, we can equivalently transform the original problem into a DUARA problem as follows
\begin{subequations}
\label{UARA}
\begin{align}
\label{eqUARA} &\min\limits_{\bm{X},\bm{Y},\bm{Z}} \text{EDT}^{\text{max}}(\bm{X},\bm{Y},\bm{Z},\bm{p}^\ast,  \bm{\eta}^\ast, \bm{f}^\ast ) \\
\label{UARA1}  &\quad \text{s.t.} \quad  \eqref{US_A1}\sim\eqref{RB_UE_e2},\eqref{A_constraints}\sim\eqref{11c}.
\end{align}
\end{subequations}
Therefore, the continuous variables and 0-1 variables in original problem \eqref{formulationProblem} can be jointly optimized by solving DUARA problem \eqref{UARA}.

\section{Proposed Iterative Parametric Convex Approximation Algorithm}
\label{SCA_section}
In this section, we first reformulate the PATACRA subproblem into a tractable form, then propose an IPCA algorithm to solve the reformulated subproblem, and finally analyze the convergence and complexity of the proposed IPCA algorithm.

\subsection{Reformulation of PATACRA Subproblem}
Given 0-1 variables $\bm{X}$, $\bm{Y}$, and $\bm{Z}$, we assume that UE $U_n$ is associated with helper $H_m$, server $S_k$ and RB $RB_l$ for notation convenience. To tackle the non-convex objective function in PATACRA subproblem \eqref{CCRA} more conveniently, we introduce an auxiliary variable $ {\phi} $ to eliminate the $\max$ operation from the objective function. Therefore, subproblem \eqref{CCRA} is transformed as
\begin{subequations}
\label{CCRA_re1}
\begin{align}
&\min\limits_{\bm{p}, \bm{\eta}, \bm{f},  {\phi}}  \phi \\
&\quad\quad \text{s.t.} \quad  \eqref{delayeq}, \eqref{energyofHelper} \sim \eqref{f_all_constraint}, \eqref{constrain_Tn}\sim\eqref{11n}, \\
\label{cons_CCRA_re1_EDT}&\quad\quad\quad\quad  \omega_n^E  E_n  + \omega_n^T T_n \leq \phi, \forall n\in \mathcal{N},
\end{align}
\end{subequations}
which is still non-convex due to the non-convexity of constraints \eqref{delayeq}, \eqref{constrain_Tn}, and \eqref{cons_CCRA_re1_EDT}.

Since \eqref{delayeq} is an equality constraint, transmit powers $p_{n}^H$ and $p_{n}^S$ can be converted to the following functions with respect to $\tau _{n}^{\text{tran}}$, $\eta_n^H$, and $\eta_n^S$, indicating that original variables $p_{n}^H$ and $p_{n}^S$ can be removed from problem \eqref{CCRA_re1}. Accordingly, the converted functions are given by
\begin{align}
\label{p1}
p_{n}^H  &\triangleq A_{n,2} g \left( \frac{{\eta_{n}^{\text{O},1} {D_n}}}{{{B}\tau _{n}^{\text{tran}}}} \right),
\end{align}
and
\begin{align}
\label{p2}
p_{n}^S  &\triangleq A_{n,1} g \left( \frac{{\eta _{n}^{\text{O},2} {D_n}}}{{{B}\tau _{n}^{\text{tran}}}} \right)-A_{n,2} g \left(\frac{{\eta _{n}^{\text{O},1} {D_n}}}{{{B}\tau _{n}^{\text{tran}}}}\right) \notag\\
&+ A_{n,2} g \left( \frac{{\eta _{n}^H {D_n} + \eta _{n}^S D_n}}{{{B}\tau _{n}^{\text{tran}}}} \right),
\end{align}
where $g(x) = 2^{x}-1$, $A_{n,1} \triangleq {(1-2\text{O}_{n,k,m,l})({1}/{{{ {g_{n,k,l}^S}}}} -  {1}/{{{ {g_{n,m,l}^H} }}})} \geq 0$, $A_{n,2} \triangleq (1-\text{O}_{n,k,m,l})/{{{{g_{n,m,l}^H} }}} + \text{O}_{n,k,m,l}/{g_{n,k,l}^S}$, $\eta_{n}^{\text{O},1} \triangleq (1-\text{O}_{n,k,m,l})\eta_n^H + \text{O}_{n,k,m,l}\eta_n^S  $, $\eta _{n}^{\text{O},2}\triangleq \text{O}_{n,k,m,l}\eta_n^H + (1-\text{O}_{n,k,m,l})\eta_n^S$, and $\tau _{n}^{\text{tran}}=\tau^{\text{tran-H}}_{n}=\tau^{\text{tran-S}}_{n}$ for notation convenience. The proof of deriving \eqref{p1} and \eqref{p2} can be referred to Appendix A in \cite{power_derivation}.
Recall that $T_n\triangleq \max \{ t_n^L, t^{\text{off-H}}_n, t^{\text{off-S}}_n \}$, which is difficult to deal with. Therefore, we introduce auxiliary variables $\bm{\beta} \triangleq \{ \beta_n| \forall n \in \mathcal{N}\}$ to eliminate $\max$ operation by approximating the upper bound of $T_n$, i.e., $\beta_n \geq T_n$, $\forall n $. Then, by substituting \eqref{p1}, \eqref{p2}, and $\bm{\beta}$ into \eqref{CCRA_re1}, we have
\begin{subequations}
\label{eqCn}
\begin{align}
\label{Cna} & \quad \quad \min\limits_{\bm{\tau}, \bm{\eta}, \bm{f}, \bm{\beta}, \phi} \quad { \phi}\\
\label{Cnb} &\quad\quad \text{s.t.}  \quad \eqref{energyofHelper} \sim \eqref{f_all_constraint},\\
\label{Cne}&\quad\quad\quad\quad  \beta_n \le T_n^{\max}, \forall n \in \mathcal{N}, \\
\label{Cnm}&\quad\quad\quad\quad   p_{n}^H  + p_{n}^S \leq P_n^{\max}, \forall n \in \mathcal{N}, \\
\label{Cnf} &\quad\quad\quad\quad   \frac{(1-\eta_n^H-\eta_n^S)D_nI_n}{f_n^L} \le {\beta_n},  \forall n \in \mathcal{N},\\
\label{Cng}&\quad\quad\quad\quad \tau^{\text{tran}}_{n} +  \tau_{n}^\text{exe-H}\le {\beta_n},  \forall n \in \mathcal{N},\\
\label{Cnh}&\quad\quad\quad\quad \tau^{\text{tran}}_{n} + \tau_{n}^\text{exe-S}\le {\beta _n},  \forall n \in \mathcal{N}, \\
\label{cons_CCRA_re2_EDT}&\quad\quad\quad\quad  \omega_n^E  E_n^\prime  + \omega_n^T \beta_n \leq \phi, \forall n\in \mathcal{N},
\end{align}
\end{subequations}
where $\bm{\tau} = \{\tau_n^{\text{tran}} |\forall n \in \mathcal{N}\}$ is the transmission delay, and $E_n^\prime$ in constraint \eqref{cons_CCRA_re2_EDT} is given by
\begin{align}
\label{eqn_dbl_x}
E_n^\prime & \triangleq  \left( {1 - \eta _{n}^H - \eta _{n}^S} \right){D_n}{I_n}\zeta {\left( {f_n^L} \right)^2} +  \tau_{n}^{\text{tran}}A_{n,1} g \left( \frac{{\eta _{n}^{\text{O},2} {D_n}}}{{{B}\tau _{n}^{\text{tran}}}} \right) \notag \\
&+ \tau _{n}^{\text{tran}} A_{n,2} g \left( \frac{{\eta _{n}^H {D_n} + \eta _{n}^S D_n}}{{{B}\tau _{n}^{\text{tran}}}} \right).
\end{align}

\textbf{Proposition 1.} Constraint \eqref{cons_CCRA_re2_EDT} is a convex constraint.
\begin{IEEEproof}
Please see Appendix A.
\end{IEEEproof}
Note that problem \eqref{eqCn} is still non-convex due to the non-convexity of constraint \eqref{Cnm} with respect to the coupled variables $\tau _{n}^{\text{tran}}$, $\eta _{n}^H$, and $\eta _{n}^S$, and the non-convenxity of constraint \eqref{Cnh} with respect to coupled variables $\eta_n^S$ and $f_{n,k}^S$. To solve this non-convex optimization problem, a general scheme is to approximate the non-convex feasible set by using successive convex approximation methods \cite{SCA_math,SCA_ZW,CL_ZW,SCA_BM}. Specifically, denote a non-convex problem as $\min Y(x)\: \text{s.t.} \: g_i(x) \leq 0, \forall i \in \{1,...,I\}$, where $g_i(x)$ is a non-convex function and $I$ is the number of constraints. The basic idea of the successive convex approximation method is that non-convex $g_i(x)$ is replaced by the upper convex and continuously differentiable function $G_i(x, y_r)$ at iteration $r$, where $y_r$ is a appropriately selected parameter\cite{SCA_math}.

However, for non-convex problem \eqref{eqCn}, since constraints \eqref{Cnm} and \eqref{Cnh} have non-convex terms $A_{n,1} g \left( \frac{{\eta _{n}^{\text{O},2} {D_n}}}{{{B}\tau _{n}^{\text{tran}}}} \right)+A_{n,2} g \left( \frac{{\eta _{n}^H {D_n} + \eta _{n}^S D_n}}{{{B}\tau _{n}^{\text{tran}}}} \right)$ and $\tau_n^{\text{tran}}+\frac{\eta_{n}^S D_n I_n}{  f_{n,k}^S}$, respectively, it is difficult to directly find their upper convex approximations. Therefore, in the next subsection, we first transform these non-convex items into the form that can be approximated by the convex approximation methods, and then propose an IPCA algorithm to solve problem \eqref{eqCn}.
\subsection{Convex Approximation Transformation and Proposed IPCA Algorithm}
To construct the upper convex approximations of the non-convex items in \eqref{Cnm} and \eqref{Cnh}, we replace the intractable non-convex items by introducing auxiliary variables, and then transform the replaced constraints into the form that can adopt the convex approximation. The transformation details of each constraint are described as follows.

\textit{For non-convex constraint \eqref{Cnm}}, it can be transformed as the following convex constraint by introducing auxiliary variables $\{ z_{n,1}, z_{n,2}| \forall n \in \mathcal{N}\}$
\begin{align}
A_{n,1}( 2^{z_{n,1}} - 1) + A_{n,2} ( 2^{z_{n,2}} - 1)  \leq P_n^{\max}, \forall n \in \mathcal{N}, \label{e15Cnm}
\end{align}
where $z_{n,1}$ and $z_{n,2}$ should satisfy
\begin{align}
{\frac{{(\text{O}_{n,k,m,l}\eta_n^H + (1-\text{O}_{n,k,m,l})\eta_n^S)}D_n}{{B \tau _{n}^{\text{tran}}}}} \le z_{n,1}, \forall n \in \mathcal{N},  \label{z1z1}
\end{align}
and
\begin{align}
 {\frac{(\eta _{n}^H+\eta _{n}^S){D_n}}{{B \tau _{n}^\text{tran}}}} \le z_{n,2}, \forall n \in \mathcal{N}. \label{z1z2}
\end{align}
To transform introduced non-convex constraints \eqref{z1z1} and \eqref{z1z2} into a form that can be approximated by the successive convex approximation method, i.e., first-order Taylor approximation\cite{SCA_ZW}, we introduce auxiliary variables $\{u_{n,1}, u_{n,2}| \forall n \in \mathcal{N} \}$, and then constraints \eqref{z1z1} and \eqref{z1z2} can be rewritten as follows
\begin{align}
z_{n,i} \tau _{n}^{\text{tran}} \geq u_{n,i}^2, \forall i \in \{1,2\}, \forall n \in \mathcal{N}, \label{e27}
\end{align}
\begin{align}
u_{n,1}^2 \geq {\frac{{(\text{O}_{n,k,m,l}\eta_n^H + (1-\text{O}_{n,k,m,l})\eta_n^S) {D_n}}}{{B}}}, \forall n \in \mathcal{N}, \label{e28z1}
\end{align}
and
\begin{align}
 u_{n,2}^2 \geq {\frac{(\eta _{n}^H+\eta _{n}^S){D_n}}{B}}, \forall n \in \mathcal{N}.\label{e28}
\end{align}
Then, non-linear constraint \eqref{e27} can be rewritten as a convex linear matrix inequality (LMI) constraint
\begin{align}
\label{eLMI}
\begin{bmatrix}
z_{n,i} & u_{n,i} \\
u_{n,i} & \tau _{n}^{\text{tran}}
\end{bmatrix} \succeq 0, \forall i \in \{1,2\}, \forall n \in \mathcal{N}.
\end{align}

For non-convex constraint \eqref{e28z1} and \eqref{e28}, by using the first-order Taylor approximation, we introduce surrogate functions $G_{n,r,i}(u_{n,i}, u_{n,i}^{(r)})= 2 u_{n,i} u_{n,i}^{(r)} - (u_{n,i}^{(r)})^2, \forall i \in \{1,2\}$ as convex lower bounds for $u_{n,i}^2$, respectively, which can be given by
\begin{align}
\label{SCA_u_all}
G_{n,r,i}(u_{n,i}, u_{n,i}^{(r)}) \leq u_{n,i}^2, \forall i \in \{1,2\},
\end{align}
where $u_{n,i}^{(r)}=u_{n,i}^{(r-1)}$ at iteration $r$.
Then, non-convex constraints \eqref{e28z1} and \eqref{e28} around points $u_{n,i}^{(r)}$ at iteration $r$ can be expressed by
\begin{align}
\frac{{{(\text{O}_{n,k,m,l}\eta_n^H + (1-\text{O}_{n,k,m,l})\eta_n^S)} {D_n}}}{B} &\leq G_{n,r,1}(u_{n,1}, u_{n,1}^{(r)}), \label{e33}
\end{align}
\begin{align}
{\frac{(\eta _{n}^H +\eta _{n}^S) {D_n}}{B}} &\leq G_{n,r,2}(u_{n,2}, u_{n,2}^{(r)}). \label{e34}
\end{align}
Note that \eqref{e33} and \eqref{e34} are convex constraints since both sides of the inequalities are affine functions.

\textit{For non-convex constraint \eqref{Cnh}}, we introduce variables $\{\xi_n| \forall n \in \mathcal{N}\}$ to tackle the non-convex term $t_{n}^{\text{exe-S}}, \forall n \in \mathcal{N}$. Therefore, \eqref{Cnh} can be transformed as
\begin{align}
\label{f_nk_constraint}
\tau^{\text{tran}}_{n} +\xi_n \le {\beta _n},  \forall n \in \mathcal{N},
\end{align}
and
\begin{align}
\label{f_nk_xi_constraint}
\frac{\eta_{n}^S D_n I_n}{ f_{n,k}^S} \leq \xi_n, \forall n \in \mathcal{N}.
\end{align}
Since \eqref{f_nk_xi_constraint} is a non-convex constraint, we can transform it as a convex constraint by introducing variables $\{u_{n,3}|\forall n \in \mathcal{N}\}$, which is given by
\begin{align}
\label{f_nk_trans1}
\frac{\xi_n  f_{n,k}^S}{D_n I_n} \geq u_{n,3}^2,\forall n \in \mathcal{N},
\end{align}
and
\begin{align}
\label{f_nk_trans2}
u_{n,3}^2 \geq \eta_n^S,\forall n \in \mathcal{N},
\end{align}
where \eqref{f_nk_trans1} can be rewritten as a convex LMI constraint
\begin{align}
\label{f_nk_trans1_LMI}
\begin{bmatrix}
\frac{\xi_n}{D_n} & u_{n,3} \\
u_{n,3} & \frac{ f_{n,k}^S}{I_n}
\end{bmatrix} \succeq 0, \forall n \in \mathcal{N}.
\end{align}

Likewise, for introduced non-convex constraint \eqref{f_nk_trans2}, by using the first-order Taylor approximation, we introduce surrogate functions $G_{n,r,3}(u_{n,3}, u_{n,3}^{(r)})= 2 u_{n,3} u_{n,3}^{(r)} - (u_{n,3}^{(r)})^2$ as convex lower bound for $u_{n,3}^2$, where $u_{n,3}^{(r)}=u_{n,3}^{(r-1)}$ at iteration $r$. Then, constraint \eqref{f_nk_trans2} can be transformed as
\begin{align}
\label{Taylor_u_3n}
 \eta_n^S \leq G_{n,r,3}(u_{n,3}, u_{n,3}^{(r)}),
\end{align}
which is a convex constraint.

Overall, replacing \eqref{e28z1}, \eqref{e28}, and \eqref{f_nk_trans2} by their corresponding approximate constraints at iteration $r$, subproblem \eqref{eqCn} can be approximately transformed into a convex subproblem shown as follows
\begin{subequations}
\label{equ_final}
\begin{align}
\label{eq3a} &\min\limits_{\bm{\tau}, \bm{\eta}, \bm{f}, \bm{\beta},\bm{z}, \bm{\xi}, \bm{u},\phi} \quad { \phi}\\
\label{3a} &\quad \text{s.t.}\quad \eqref{Cnb},\eqref{Cne},\eqref{Cnf},\eqref{Cng},\eqref{cons_CCRA_re2_EDT},\\
&\quad\quad\quad\eqref{e15Cnm},\eqref{eLMI}, \eqref{e33}\sim\eqref{f_nk_constraint},\eqref{f_nk_trans1_LMI},\eqref{Taylor_u_3n},
\end{align}
\end{subequations}
where $\bm{z} = \{z_{n,i}|\forall i \in \{1,2\},\forall n \in \mathcal{N}\}$, $\bm{\xi} = \{\xi_{n}|\forall n \in \mathcal{N}\}$, and $\bm{u} = \{u_{n,i}| \forall i \in \{1,2,3\}, \forall n \in \mathcal{N}\}$ are the introduced auxiliary variables.

It can be seen that subproblem \eqref{equ_final} is a convex optimization problem, which can be solved by the convex solver, i.e., CVX toolbox\cite{boyd2004convex,SCA_ZW,CL_ZW,SCA_BM}, at iteration $r$. Nevertheless, since $\tau_{n}^{\text{tran}}A_{n,1} g \left( \frac{{\eta _{n}^{\text{O},2} {D_n}}}{{{B}\tau _{n}^{\text{tran}}}} \right)$ and $\tau _{n}^{\text{tran}} A_{n,2} g \left( \frac{{\eta _{n}^H {D_n} + \eta _{n}^S D_n}}{{{B}\tau _{n}^{\text{tran}}}} \right)$ in convex constraint \eqref{cons_CCRA_re2_EDT} are perspective functions with structure $y2^{x/y}$, they will be replaced and solved by using convex function tool $\{x\ln(2), y, z\} == exponential(1)$ of the CVX toolbox \cite{IoTMEC1}. Please refer to \cite{CVX_tool_book} for the details of this substitution.

With the above transformations and convex approximations, we propose an IPCA algorithm to directly solve subproblem \eqref{eqCn} by iteratively solving a sequence of approximated convex problem.
Specifically, by solving approximated convex subproblem \eqref{equ_final} at each iteration, subproblem \eqref{eqCn} can be approximately solved until convergence, where the parameters introduced for convex approximation of problem \eqref{equ_final} are dependent on the solution of the previous iteration. The procedure of the proposed IPCA algorithm is described in Algorithm 1, where $\epsilon$ is a small positive value to control the accuracy of the proposed IPCA algorithm.
\begin{table}[h]
    \small
    \begin{tabular}{p{240pt}}
    \toprule
    \textbf{Algorithm 1:} The Proposed IPCA Algorithm\\
    \midrule
    1: \textbf{Initialization}: Initialize the initial feasible points $\bm{u}^{(0)}$, $\phi^{(0)}$, \\
        \quad  and set $r = 1$;\\
    \hangafter 1 
    \hangindent 2em 
    2: \textbf{While} $|\phi^{r}-\phi^{r-1}| \geq \epsilon$ \textbf{do}\\
    3:          \quad Solve problem \eqref{equ_final}  and get $\bm{\tau}^{(r)}, \bm{\eta}^{(r)}, \bm{f}^{(r)}, \bm{\beta}^{(r)},\bm{z}^{(r)}, \bm{\xi}^{(r)}$, \\
                \quad \quad  $\bm{u}^{(r)}$, and $\phi^{(r)}$;\\
    4:          \quad Set $r = r+1$; \\

    5:          \quad Update $ \bm{u}^{(r)} = \bm{u}^{(r-1)}$;\\
    6: \textbf{end while}\\
    7: \textbf{Output}: $\bm{\tau}^{\ast}, \bm{\eta}^{\ast}, \bm{f}^{\ast}$ and $\phi^{\ast}$. \\
    9: Calculate $\bm{p}^\ast$ by utilizing \eqref{p1} and \eqref{p2}.\\
    \bottomrule
    \end{tabular}
    \vspace{-1em}
\end{table}
\subsection{Convergence and Complexity Analysis}
To prove the convergence of the proposed IPCA algorithm, the proposition is provided as follows.

\textbf{Proposition 2.} A sequence of non-increasing objective values are generated by the proposed IPCA algorithm and finally converges.
\begin{IEEEproof}
First, recall that $\frac{{\eta _{n}^S {D_n}}}{B} \leq G_{n,r,1}(u_{n,1}, u_{n,1}^{(r)})\leq u_{n,1}^2$, ${\frac{(\eta _{n}^H +\eta _{n}^S) {D_n}}{B}} \leq G_{n,r,2}(u_{n,2}, u_{n,2}^{(r)})\leq u_{n,2}^2$, and $ \eta_n^S \leq G_{n,r,3}(u_{n,3}, u_{n,3}^{(r)}) \leq u_{n,3}^2$. It can be concluded that the feasible point of subproblem \eqref{equ_final} at iteration $r$ is also feasible for subproblem \eqref{eqCn}.

Next, we will prove that a sequence of non-increasing objective values can be generated.
Recall that for iteration $r$, convex constraints \eqref{e33}, \eqref{e34}, and \eqref{Taylor_u_3n} are transformed from \eqref{e28z1}, \eqref{e28}, and \eqref{f_nk_trans2}. For constraint \eqref{e28z1}, we have $\frac{{\eta _{n}^S {D_n}}}{B} \leq G_{n,r,1}(u_{n,1}, u_{n,1}^{(r)})$. Then, assume that the optimal solution to subproblem \eqref{equ_final} at iteration $r$ is $\{\bm{\tau}^{(r)\ast}, \bm{\eta}^{(r)\ast}, \bm{f}^{(r)\ast}, \bm{\beta}^{(r)\ast},\bm{z}^{(r)\ast}, \bm{\xi}^{(r)\ast}, \bm{u}^{(r)\ast},\phi^{(r)\ast}\}$.
By replacing $\bm{u}^{(r+1)}$ with the solution obtained from iteration $r$, i.e., $\bm{u}^{(r+1)}=\bm{u}^{(r)\ast} $, we have that
\begin{align}
 G_{n,r,1}(u_{n,1}, u_{n,1}^{(r+1)}) =  2 u_{n,1} u_{n,1}^{(r)\ast} - (u_{n,1}^{(r)\ast})^2 \geq \frac{{\eta _{n}^S {D_n}}}{B},
\end{align}
which indicates that the solution of subproblem \eqref{equ_final} at iteration $r$ satisfies constraint \eqref{e28z1} at the next iteration. Following the same rationale, we can prove that the solution obtained at iteration $r$ also satisfies constraints \eqref{e33} and \eqref{e34} at iteration $r+1$. Therefore, the solution obtained at iteration $r$ is a feasible point for subproblem \eqref{equ_final} at next iteration $r+1$, indicating that the objective value is monotonically non-increasing with the iteration index.

In a summary, it can be concluded that a sequence of feasible solutions can be obtained with non-increasing objective values. Additionally, since the feasible set of problem \eqref{eqCn} is nonempty and compact, the sequence of non-increasing objective values is lower-bounded. Therefore, the proposed IPCA algorithm will converge to a stationary point.
\end{IEEEproof}

Since subproblem \eqref{equ_final} is solved by the CVX toolbox using interior-point method \cite{boyd2004convex,SCA_ZW,CL_ZW,SCA_BM}, the complexity of Algorithm 1 is $\mathcal{O}(\delta)=\mathcal{O}(l_1 \frac{\log (S/(\epsilon t ))}{\log \theta})$ \cite{boyd2004convex},
where $\epsilon$ denotes a guaranteed specified accuracy, $t$ is the initial centering point, $\theta$ denotes the updating step size to increase $t$, $S$ is the number of constraints, and $l_1$ is the iteration number of Algorithm 1.
\section{Proposed Four-sided UE-RB-helper-server Matching Algorithm}
\label{3DMatching}
In this section, we first model DUARA problem \eqref{UARA} as a four-sided matching problem. Then, using the objective value provided by the IPCA algorithm to determine the utility, we propose an FS-URHSM algorithm to solve the DUARA problem. Finally, we analyze the stability and complexity of the proposed FS-URHSM algorithm.
\subsection{Modeling of Four-Sided Matching}
Recall that DUARA problem \eqref{UARA} only contains discrete 0-1 variables $\bm{X}$, $\bm{Y}$, and $\bm{Z}$, which is shown to be NP-hard and can not be directly solved with polynomial complexity. However, if the exhaustive search algorithm is used to find the optimal solution by traversing all possible combinations of variables $\bm{X}$, $\bm{Y}$, and $\bm{Z}$, it will suffer an exponential complexity of $\mathcal{O}(\delta 2^{NM+NK+NL})$. Fortunately, as a promising approach for dynamic resource allocation and network management\cite{matchingTheory_A,Matching_Zonghshu}, \emph{matching theory} can offer a near-optimal solution with much lower complexity than the exhaustive search. In specific, matching theory utilizes preferences to describe the dynamic relationship between different sets of players, improves players' preferences by using deferred-acceptance/swap operations, and finally achieves a stable matching that no deferred-acceptance/swap operation can improve the players' utilities/preferences.
Conventional matching theory mainly focused on the two-sided \cite{Multi_NOMAMEC_SM} or three-sided matching \cite{3D_TMC_cyclic,huang2010circular,3DMatching_Zhang} with cyclic/mixed preferences, e.g., one-to-one two-sided marriage problem introduced by Gale and Shapley \cite{matchingTheory_A}, many-to-one two-sided college admission problem, and three-sided kidney transplant problem \cite{huang2010circular}. However, since DUARA problem \eqref{UARA} contains 0-1 variables $\bm{X}$, $\bm{Y}$, and $\bm{Z}$, it can not be modeled as a conventional two/three-sided matching problem, indicating that a new four-sided matching problem should be designed.

Inspired by above observations, we model DUARA problem \eqref{UARA} as a four-sided matching problem among UEs, RBs, MEC helpers, and servers in Definition 1. Before this definition, we give the following notations according to the conventional two/three-sided matching\cite{matchingTheory_A} to better describe the modeling of four-sided matching.
\begin{itemize}
\item \emph{Matching triple}: If helper $H_m$, MEC server $S_k$, and RB $RB_l$ are matched to each other, they can be defined as a matching triple, denoted as $(H_m, S_k, RB_l )$ $\in$ $\mathcal{M} \times \mathcal{K} \times \mathcal{L}$. Likewise, $(U_n, S_k, RB_l)$ $\in$ $\mathcal{N} \times \mathcal{K} \times \mathcal{L}$, $ (U_n, H_m, S_k)$ $\in$ $\mathcal{N} \times \mathcal{M} \times \mathcal{K}$, and $(U_n, H_m, S_k)$ $\in$ $\mathcal{N} \times \mathcal{M} \times \mathcal{K}$ are also matching triples, respectively.
\item \emph{Matching unit}: If UE $U_n$ matches to a matching triple $(H_m, S_k, RB_l ) \in \mathcal{M} \times \mathcal{K} \times \mathcal{L}$, it indicates that helper $H_m$, MEC server $S_k$, and RB $RB_l$ are matched to each other, which can be viewed as a matching unit defined as $\mu_n \triangleq (U_n, H_m, S_k, RB_l)$.
\end{itemize}

With above introduced notations, DUARA problem \eqref{UARA} with only discrete 0-1 variables can be modeled as the following four-sided matching model.

\textbf{Definition 1}: \textit{(Four-sided Matching Model)}
\emph{Four-sided matching} $\Psi$ represents the mapping relationship among UEs, RBs, MEC helpers, and servers, which satisfies the following conditions:
\begin{enumerate}
\item[{1)}] $\Psi (U_n) = (H_m, S_k, RB_l ) \in \mathcal{M} \times \mathcal{K} \times \mathcal{L}$, $|\Psi (U_n)| = \{1,1,1\}$, $\forall n \in \mathcal{N}$;
\item[{2)}] $\Psi (H_m) = (U_n, S_k, RB_l) \in \mathcal{N} \times \mathcal{K} \times \mathcal{L}$, $ |\Psi (H_m)| = \{1,1,1\}$, or MEC helper $H_m$ is not matched to any UE, server and RB, $\forall m \in \mathcal{M}$;
\item[{3)}] $\Psi (S_k) = (U_n, H_m, RB_l) \in \mathcal{N} \times \mathcal{M} \times \mathcal{L}$, $ |\Psi (S_k)| = \{N_k^{\max},N_k^{\max},N_k^{\max}\}$, or server $S_k$ is not matched to any UE, helper and RB, $\forall k \in \mathcal{K}$;
\item[{4)}] $\Psi (RB_l) = (U_n, H_m, S_k) \in \mathcal{N} \times \mathcal{M} \times \mathcal{K}$, $|\Psi (RB_l)| = \{1,1,1\}$, or RB $RB_l$ is not matched to any UE, MEC helper and server, $\forall l \in \mathcal{L}$;
\item[{5)}] $\Psi (U_n) = (H_m, S_k, RB_l ) \Leftrightarrow \Psi (H_m) = (U_n, S_k, RB_l) \Leftrightarrow \Psi (S_k) = (U_n, H_m, RB_l) \Leftrightarrow \Psi (RB_l) = (U_n, H_m, S_k)$, $\forall n \in \mathcal{N}$;
\end{enumerate}
where the positive integers in right hand of equation $|\Psi (\cdot)| =\{1,1,1\}$ and $|\Psi (S_k)| =\{N_k^{\max},N_k^{\max},N_k^{\max}\}$ denote the number of matched players, respectively. Conditions 1)$\sim$5) can be explained as follows. Condition 1) indicates that each UE is matched to one helper, one server, and one RB. Condition 2) indicates that each MEC helper can be matched to at most one UE, one server, and one RB. Condition 3) means that each server is matched to at most $N_n^{\max}$ UEs/helpers/RBs. Condition 4) means that each RB is matched to at most one UE, one helper, and one server. Condition 5) indicates that UE $U_n$, helper $H_m$, server $S_k$, and RB $RB_l$ are matched to each other.

Therefore, with the above definition of matching $\Psi$, DUARA problem \eqref{UARA} can be equivalently viewed as finding the best four-sided matching to minimize the utility function, which is defined as
\begin{align}
\label{UF}
U(\Psi) &\triangleq \text{EDT}^{\text{max}} [ \bm{X}(\Psi),\bm{Y}(\Psi),\bm{Z}(\Psi),\bm{p}^\ast,  \bm{\eta}^\ast, \bm{f}^\ast ],
\end{align}
where $\bm{X}(\Psi)$, $\bm{Y}(\Psi)$ and $\bm{Z}(\Psi)$ denote the user association and RB assignment based on $\Psi$. The elements in $\{\bm{X}(\Psi), \bm{Y}(\Psi), \bm{Z}(\Psi)\}$ can be characterized by
\begin{align}
\{x_{n,k},y_{n,m}, z_{n,l}\} =
\begin{cases}
\{1,1,1\}, & \text{for} \: \mu_n \in \Psi, \forall n \in \mathcal{N};\\
\{0,0,0\}, & \text{otherwise}.
\end{cases}
\end{align}
\subsection{Proposed FS-URHSM Algorithm}
In conventional two/three-sided matching theory, the deferred-acceptance/swap algorithms with cyclic/mixed preference are designed to find the best feasible matching \cite{3D_TMC_cyclic,huang2010circular,3DMatching_Zhang}. For this four-sided matching problem, if the deferred-acceptance/swap algorithms are directly utilized, the players in each set need to establish the cyclic/mixed preference of the players in the other three sets. However, since each UE's preference of server/helper/RB may change with the matching conditions of other UEs matched to the same server, it is difficult to accurately describe this dynamic preferences with using the cyclic/mixed preferences. In addition, if the deferred-acceptance/swap algorithms are directly used to find the best four-sided matching, the feasibility of the four-sided matching may be destroyed and some potential matchings with better performance will be ignored. Specifically, due to the dynamic nature of preferences in the four-sided matching problem, some deferred-acceptance operations may be repeatedly performed, which may lead to the failure of the matching process and result in the infeasible matching. Moreover, conventional swap operations are suitable for players who are already matched to each other \cite{Multi_NOMAMEC_Zhu}, ignoring some unmatched players with better performance. Therefore, we first define the \emph{matching utility} of a feasible matching $\Psi$ with the solution obtained from PATACRA subproblem \eqref{CCRA}, and then propose the FS-URHSM algorithm with the newly designed operations as follows.

To find the best four-sided matching, we use \emph{matching utility} to describe the competition behavior and decision process among UEs, helpers, servers, and RBs, which is also from utility function \eqref{UF}.
Specifically, for matching $\Psi$, the matching utility is given by
\begin{align}
\label{UFfinal}
U(\Psi) &\triangleq \phi^{(\ast)},
\end{align}
where $\phi^{(\ast)}$ can be obtained by our proposed IPCA algorithm in Algorithm 1.

With the matching utility defined above, we propose a novel FS-URHSM algorithm to find out the best matching by repeatedly performing operations in Definitions 2$\sim$5 until there is no defined blocking matching.

\textbf{Definition 2}: \textit{(Sequential Swap (SS) Operation and SS Matching)}
Suppose $\mu_1$ and $\mu _2$ are two different matching units of matching $\Psi$, expressed as
\begin{align}
\mu _1 \triangleq (U_n, H_m, S_k, RB_l),
\end{align}
and
\begin{align}
\mu _2 \triangleq  (U_{n^\prime}, H_{m^\prime}, S_{k^\prime}, RB_{l^\prime}),
\end{align}
where $n \neq n^\prime, m \neq m^\prime, k \neq k^\prime$ and $l \neq l^\prime$.
Assume that two players $a \in \mu_1$ and $a' \in \mu _2$ belongs to the same set $\mathcal{N}$, $\mathcal{M}$, $\mathcal{K}$ or $\mathcal{L}$, respectively. The sequential swap operation over $a$ and $a'$ means that players $a$ and $a'$ will exchange their currently matched players $\Psi(a)$ and $\Psi(a')$. Moreover, the matching conditions of matching $\Psi$ stay the same except for matching units $\mu_1$  and $\mu _2$. After the sequential swap operation, we have the following matching
\begin{align}
\Psi^{\text{SS}}_{a,a'} \triangleq & \Psi  \setminus \{ (a, \Psi(a)), (a',\Psi(a')) \} \notag \\
                                      &\cup \{ (a, \Psi(a')), (a',\Psi(a)) \},
\end{align}
called \emph{SS matching} over $\{(a,\Psi(a)),(a',\Psi(a')\}$.

\textbf{Definition 3}: \textit{(SS-blocking Matching)}
In a feasible matching $\Psi$,
matching $\Psi^{\text{SS}}_{a,a'}$ can be a SS-blocking matching if and only if
\begin{equation}
U(\Psi^{\text{SS}}_{a,a'}) < U(\Psi).
\end{equation}
Definition 3 indicates that the SS operation will be performed if and only if the utility of the newly generated SS matching will be reduced after swapping.

Recall that when $M >N$, $\sum_{k=1}^K N_k^{\max} > N$ or $K >N$, not all MEC helpers/servers/RBs can be fully matched to other players in the opposite sets. Therefore, the unmatched MEC helpers/RBs will not participate in the SS operation defined above, which may lead to a local optimal matching and cause a large gap with the optimal solution. To tackle such a challenge issue in scenarios where $M >N$, $\sum_{k=1}^K N_k^{\max} > N$ or $K >N$, we have following definitions.

\textbf{Definition 4}: \textit{(Leaving and Joining-in (LJ) Operation and LJ Matching)}
From the perspective of MEC helpers, the set of matched and unmatched MEC helpers are defined as ${H}_{matched}$ and $H_{unmatched}$, respectively.
The LJ operation over $\{(H_m, H_{m^\prime})\}$ means that a matched MEC helper $H_m$ leaves out its current matching players $\Psi(H_m)$ and a new unmatched MEC helper $H_{m'}$ joins in the same matching players $\Psi(H_m)$ while keeping other conditions unchanged. The LJ matching $\Psi^{\text{LJ}}_{m,m'}$ generated after this operation can be expressed as
\begin{align}
\Psi^{\text{LJ}}_{m,m'} \triangleq \Psi \setminus \{(m,\Psi(m))\} \cup \{(m^\prime, \Psi(m))\},
\end{align}
where $m \in H_{matched}$, $m^\prime \in H_{unmatched}$, and $\Psi^{\text{LJ}}_{m,m^\prime}$ is called \emph{LJ matching} over $\{(H_m, H_{m^\prime})\}$.

\textbf{Definition 5}: \textit{(LJ-blocking Matching)}
In a feasible matching $\Psi$, $\Psi^{\text{LJ}}_{m,m^\prime }$ can be a LJ-blocking matching if and only if
\begin{align}
U(\Psi^{\text{LJ}}_{m,m^\prime }) < U(\Psi),
\end{align}
where $m \in H_{matched}$ and $m^\prime \in H_{unmatched}$.

Following the same idea, 
define the set of matched and unmatched RBs as ${RB}_{matched}$ and $RB_{unmatched}$, we can define the LJ matching over $\{ (RB_l, RB_{l^\prime })\}$ as
\begin{align}
\Psi^{\text{LJ}}_{l,l^\prime} \triangleq \Psi \setminus \{(l,\Psi(l))\} \cup \{(l^\prime, \Psi(l))\},
\end{align}
where $l \in RB_{matched}$ and $l^\prime \in RB_{unmatched}$.
Then, $\Psi^{\text{LJ}}_{l,l^\prime}$ can be a LJ-blocking matching over $\{ (RB_l, RB_{l^\prime})\}$ if and only if
\begin{align}
U(\Psi^{\text{LJ}}_{l,l^\prime}) < U(\Psi).
\end{align}

Note that when $\sum_{k=1}^K N_k^{\max} > N$, the set of matched and not fully matched servers are defined as ${S}_{matched}$ and $S_{unmatched}$, respectively. Denote $NS_{unmatched}$ as the number of UEs that have been matched by server $S_k$ ($\in S_{unmatched}$). The LJ matching over $\{ (S_k, S_{k^\prime })\}$ is defined as
\begin{align}
\Psi^{\text{LJ}}_{k,k^\prime} \triangleq \Psi \setminus \{(k,\Psi(k))\} \cup \{(k^\prime, \Psi(k))\},
\end{align}
where $k \in S_{matched}$ and $k^\prime \in S_{unmatched}$.
Then, $\Psi^{\text{LJ}}_{k,k^\prime}$ can be a LJ-blocking matching over $\{ (S_k, S_{k'})\}$ if and only if
\begin{align}
U(\Psi^{\text{LJ}}_{k,k'}) < U(\Psi).
\end{align}

Definition 5 indicates that the LJ operation will be performed if and only if the preference utility of the generated matching will be reduced after the leaving and joining in.

With the above Definitions 2$\sim$5, the procedure of the proposed low-complexity FS-URHSM algorithm\footnote{Although we focus on the cases where $M \geq N$ and $L \geq N$, our work can also be extended to case $M < N$ or $L < N$. For instance, when $M < N$, we can add a number of dumb helpers to ensure the total number of actual and dumb helpers is no less than $N$. Then, by using the proposed FS-URHSM algorithm, the user association, RB assignment, power
allocation, task assignment and computation resource allocation can be obtained. Note that dumb helpers are aware of the surrounding environment, but unable to communicate with any other users in the network \cite{ref_dumbuser}. When UE $U_n$ is matched to a dumb helper, the corresponding task splitting coefficient is set as 0.} is summarized in Algorithm 2. In each inner ``Repeat-Until'' iteration, by finding the SS-blocking/LJ-blocking matchings, the corresponding SS and LJ operations will be repeatedly performed to minimize the utility function. Finally, a stable matching will be achieved after several outer ``Repeat-Until'' iterations.
\begin{table}[h]
\small
    \centering
    \label{Al4}
    \begin{tabular}{p{240pt}}
    \toprule
    \textbf{Algorithm 2:} The Proposed FS-URHSM Algorithm  \\
    \midrule
    1: \textbf{Initialization}: Initialize a four-sided feasible matching $\Psi$, and \\
         \quad generate the unmatched and unmatched sets according to $\Psi$;\\
    \hangafter 1 
    \hangindent 2em 
    2: \textbf{Repeat} \\
    3:  \textbf{Stage-1 Sequential Swap Operations:} \\
    4:      \quad \textbf{Repeat} \\
    5:          \quad \quad \textbf{for} $\mu _i \in \Psi$ finds $\forall \mu_j \in \Psi \setminus \mu_i$ \textbf{do}\\

    6:                \quad \quad \quad \quad $\forall a \in \mu _i$, $a^\prime \in \mu _j$, calculate $U(\Psi^{\text{SS}}_{a,a'})$; \\
    7:                \quad\quad \quad \quad \quad \textbf{if} $\Psi^{\text{SS}}_{a,a'} $ is a SS-blocking matching \textbf{then}\\
    8:                \quad\quad \quad \quad \quad \quad Update matching $\Psi=\Psi^{\text{SS}}_{a,a'} $;\\
    9:                \quad\quad \quad \quad \quad \textbf{end if}

    10:          \quad\quad \textbf{end for} \\
    11: \quad\textbf{Until} There is no SS-blocking matching; \\
    12: \textbf{Stage-2 Leaving and Joining-in Operations:} \\
    13:   \quad\textbf{if} $M >N$ \textbf{then}\\
    14:    \quad \quad \textbf{Repeat} \\
    15:        \quad \quad \quad \textbf{for} $   m \in H_{matched}$ finds $ \forall m' \in H_{unmatched}$ \textbf{do} \\
    16:                      \quad \quad \quad \quad \quad  \textbf{if} $\Psi^{\text{LJ}}_{m,m'}$ is a LJ-blocking matching \textbf{then}\\
    17:                      \quad \quad \quad \quad  \quad \quad Update matching $\Psi=\Psi^{\text{LJ}}_{m,m'}$, $H_{unmatched}$ \\
                              \quad\quad\quad \quad  \quad \quad \quad \:  and $H_{matched}$; \\
    18:                      \quad \quad \quad  \quad \quad \textbf{end if} \\
    19:        \quad  \quad \quad \textbf{end for} \\
    20:  \quad \quad \textbf{Until} There is no LJ-blocking matching; \\
    21:       \quad\textbf{end if }  \\
    22:   \quad\textbf{if} $L >N$ \textbf{then}\\ 
    23: \quad \quad \textbf{Repeat}  \\
    24:        \quad \quad \quad \textbf{for} $  l \in RB_{matched}$ finds $ \forall l' \in RB_{unmatched}$ \textbf{do}\\
    25:                      \quad \quad \quad \quad \quad  \textbf{if} $\Psi^{\text{LJ}}_{l,l^\prime}$ is a LJ-blocking matching \textbf{then}\\
    26:                      \quad \quad \quad \quad  \quad \quad Update matching $\Psi=\Psi^{\text{LJ}}_{l,l^\prime}$, $RB_{unmatched}$ \\
                              \quad\quad\quad \quad  \quad \quad \quad \:  and $RB_{matched}$; \\
    27:                      \quad \quad \quad  \quad \quad \textbf{end if} \\
    28:        \quad  \quad \quad \textbf{end for} \\
    29:  \quad \quad \textbf{Until} There is no LJ-blocking matching; \\
    30:  \quad \textbf{end if }  \\
    31:   \quad\textbf{if} $\sum_{k=1}^K N_k^{\max} > N$ \textbf{then}\\
    32: \quad \quad \textbf{Repeat}  \\
    33:        \quad \quad \quad \textbf{for} $  k \in S_{matched}$ finds $ \forall k' \in S_{unmatched}$ \textbf{do}\\
    34:                      \quad \quad \quad \quad \quad  \textbf{if} $\Psi^{\text{LJ}}_{k,k'}$ is a LJ-blocking matching \textbf{then}\\
    35:                      \quad \quad \quad \quad  \quad \quad Update matching $\Psi=\Psi^{\text{LJ}}_{k,k'}$, $S_{unmatched}$ \\
                              \quad\quad\quad \quad  \quad \quad \quad \:  and $S_{matched}$; \\
    36:                      \quad \quad \quad  \quad \quad \textbf{end if} \\
    37:        \quad  \quad \quad \textbf{end for} \\
    38:  \quad \quad \textbf{Until} There is no LJ-blocking matching; \\
    39:  \quad \textbf{end if }  \\
    40: \textbf{Until} There is no blocking matching; \\
    41: \textbf{Output}: A stable matching $\Psi$.\\
    \bottomrule
    \end{tabular}
    \vspace{-1em}
\end{table}
\subsection{Convergence, Stability and Complexity }
\subsubsection{Convergence and stability}

The stability and convergence of the proposed FS-URHSM algorithm are analyzed as follows.

\textbf{Proposition 2.}
The proposed FS-URHSM algorithm converges to a stable matching $\Psi^*$ with local optimal solution after a limited number of SS/LJ operations.
\begin{IEEEproof}
Assume that $\Psi'^*$ is an unstable four-sided matching obtained by a certain iteration. Hence, it means that there is at least one SS-blocking/LJ-blocking matching after a certain iteration.
More specifically, if $(a,\Psi(a))$ and $(a',\Psi(a'))$ satisfy the swap condition, players $a$ and $a'$ will exchange their current matching players, and the matching result $\Psi'^*$ will be updated. Likewise, if there exists a LJ-blocking matching, the LJ operations will be performed to reduce the utility.
Therefore, the utility will keep monotonically decreasing.
On the other hand, note that the number of UEs, MEC helpers, servers, and RBs are limited. Therefore, the sample space of matching is finite, which means that the number of the SS/LJ operations is finite and the utility is lower bounded \cite{Multi_NOMAMEC_SM,Matching_Y}. Recall that the utility keeps monotonically decreasing, it can be concluded that the proposed FS-URHSM algorithm will eventually converge to a stable matching after a limited number of SS/LJ operations. The proof is completed.
\end{IEEEproof}

\subsubsection{Complexity}
In the following, we analyze the complexity of the proposed FS-URHSM algorithm.

\textbf{Proposition 3.} The overall worst-case complexity of the proposed FS-URHSM algorithm is $\mathcal{O}(\Delta\delta(4\delta_1 N^2+\delta_2 N(M-N)+\delta_3 N(L-N)+\delta_4 N(K-1)))$, where $\Delta$ denotes the number of the outer iteration and $\delta$ denotes the complexity of our proposed IPCA algorithm.

\begin{IEEEproof}
For the sequential swap operations in Algorithm 2, recall that each player in the matching unit $\mu _i \in \Psi$ ($\forall i \in \{1,\ldots,N\}$) searches the corresponding player and then decides whether to carry out the sequential swap operation. Hence, the worst-case complexity of the sequential swap operations is $\mathcal{O}(4\delta_1 N^2)$, where $\delta_1$ is the number of inner ``Repeat-Until'' iterations in Stage-1.

For the leaving and joining-in operations in Stage-2 of Algorithm 2, the worst-case computation complexity comes from finding the possible LJ-blocking matching during each inner ``Repeat-Until'' iteration, which can be denoted by $\mathcal{O}(\delta_2 N(M-N)+\delta_3 N(L-N)+\delta_4 N(K-1))$, where $\delta_2$, $\delta_3$ and $\delta_4$ denote the number of inner iterations when $M>2N$, $K>N$ and $\sum_{k=1}^K N_k^{\max} > N$, respectively.
Therefore, the overall worst-case complexity of Algorithm 2 is $\mathcal{O}(\Delta\delta(4\delta_1 N^2+\delta_2 N(M-N)+\delta_3 N(L-N)+\delta_4 N(K-1)))$. The proof is completed.
\end{IEEEproof}
\section{Simulation}
\label{section_simulation}
In this section, numerical results are provided to evaluate the performance of our proposed algorithms. In the simulation setup, all the UEs/helpers/MEC servers are randomly generated within a disc centered at $(0,0)$ with the radius of 500\cite{FF_TCOM_imCSI}.
The bandwidth of each RB is set to $B_k = 1$ MHz, $\forall k \in \{1,...,K\}$, and the noise power is $\sigma^2 = -174$ dBm/Hz \cite{NOMAMEC_FF_TCOM}. The channel gain between nodes $i$ and $j$ over RB $k$ follows the exponential distribution with mean $ 1/[1+(d_{i,j}/d_0)^\alpha]$, 
where $d_{i,j}$ is the distance between nodes $i$ and $j$, $d_0$ is the reference distance, and $\alpha$ is the path loss exponent. According to the standard parameters for urban cellular networks \cite{urban}, we set $d_0 = 10$ and $\alpha = 4.7$. The maximum transmit power of each UE is set to $P_{\max} = 28$ dBm. For the task execution, we consider the computation intensity for computing 1-bit data is $I_n = 10^3$ cycles/bit and the delay tolerance $T_n^{\max}$ of each UE is $0.9$ s. For the computation of energy consumption, the effective capacitance coefficient is set to $\varsigma = 10^{-29}$ \cite{CL_para}. To reflect the heterogeneity of computation capability of the MEC servers, MEC helpers, and UEs, the computation frequencies of MEC servers are generated from uniform distribution $[20, 25]$ in Gcycles/s, helpers are generated from uniform distribution $[15, 20]$ in Gcycles/s, and the computation frequencies of UEs are generated from uniform distribution $[2, 8]$ in Gcycles/s, respectively.
Considering that the MEC helpers are always energy consumption limited, the maximal energy consumption of each helper is randomly generated from range $[0.8, 1]$ Joule. Moreover, the accuracy parameter of the IPCA algorithm is set to $\epsilon = 10^{-6}$.
\subsection{Feasibility of Proposed Algorithms}
\begin{figure}[!t]
\centering{\includegraphics[scale=0.5]{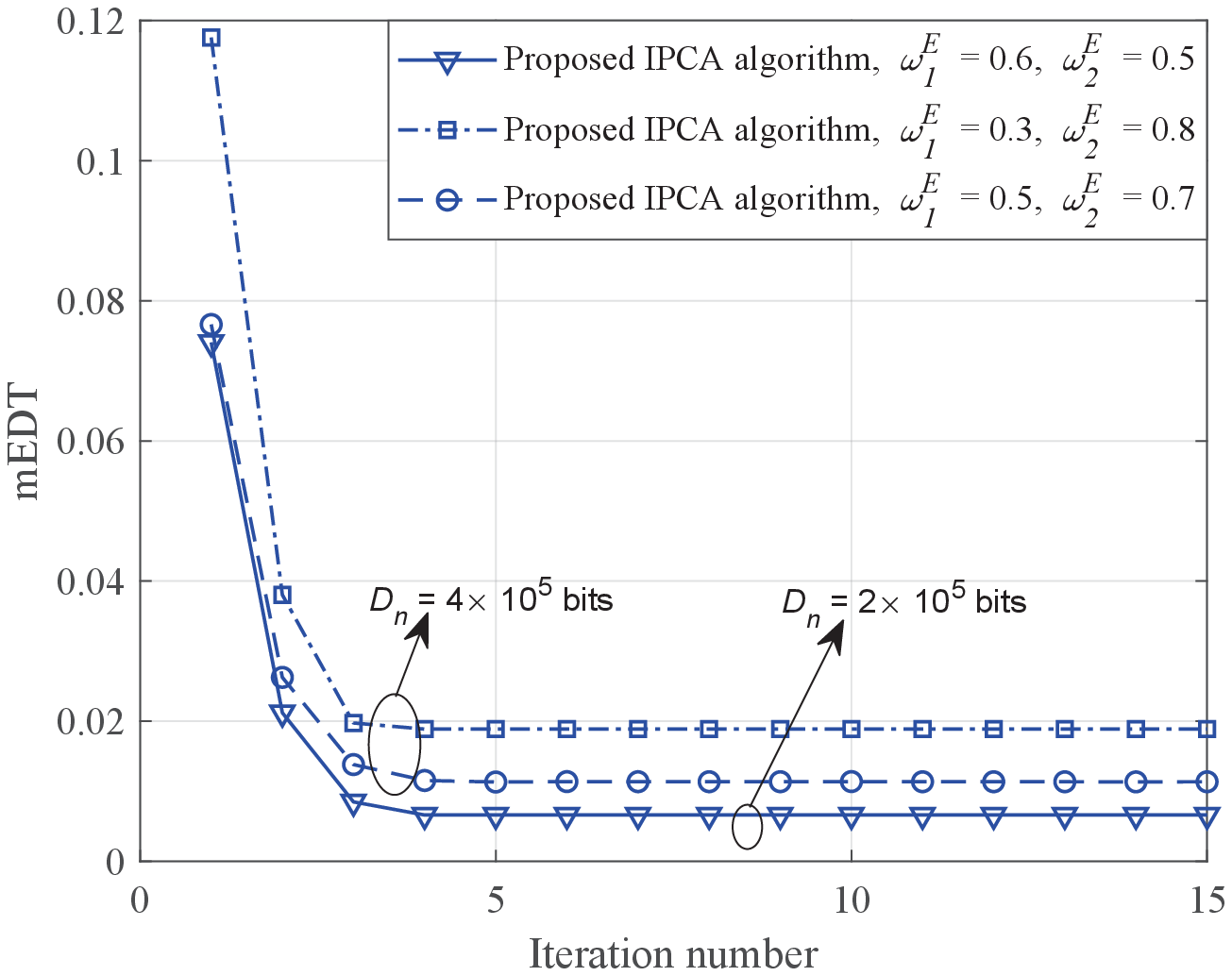}}
\caption{Convergence behavior of the proposed IPCA algorithm, where $N=K=M=L=2$ and $\omega_{n}^T = 1-  \omega_{n}^E$. }
\label{convergence1}
\end{figure}
Fig. \ref{convergence1} demonstrates the convergence behavior of the proposed IPCA algorithm. It can be observed that the proposed IPCA algorithm converges within about $5$ iterations and finally converges to a stationary point. On the other hand, it can also be seen that even with different combinations of data size and weight factors, the convergence requires nearly the same number of iterations. This observation demonstrates the convergence of the IPCA algorithm is insensitive to the values of data size and weight factors.
\begin{figure}[!t]
\centering{\includegraphics[scale=0.5]{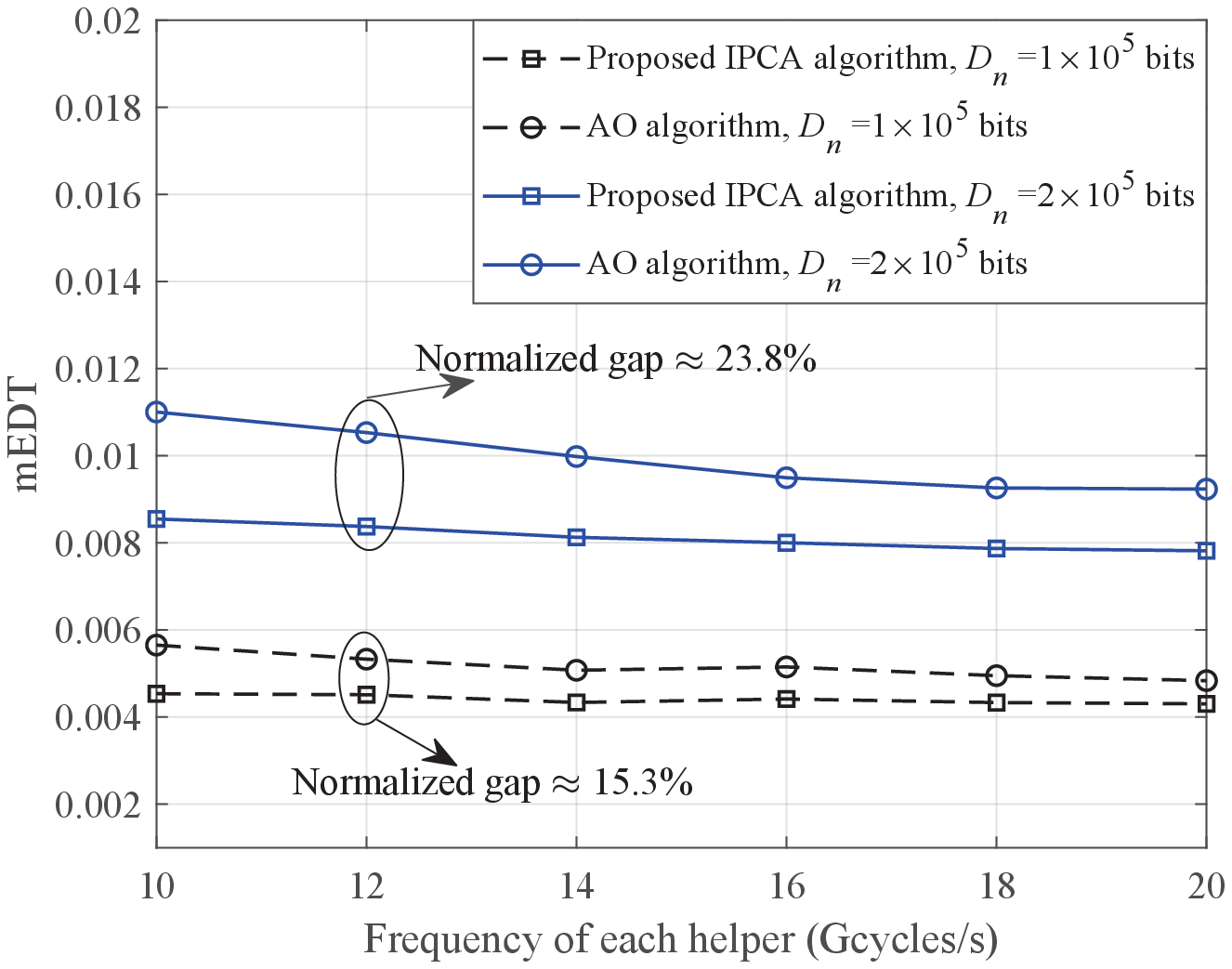}}
\caption{The mEDT comparisons of the proposed IPCA algorithm and the AO algorithm, where $N=K=M=L=2$ and $\omega_{n}^T = 1-  \omega_{n}^E$. }
\label{AO_comp}
\end{figure}

To demonstrate the performance gain of our proposed IPCA algorithm for solving non-convex problem \eqref{eqCn}, Fig. \ref{AO_comp} compares the mEDT gap between the proposed IPCA algorithm and the alternative optimization (AO) algorithm \cite{ZZ_CL_AO,ZZ_CL}. The basic idea of the AO algorithm is that it first divides the variables of \eqref{eqCn} into two blocks $\{ \bm{f}\}$ and $\{\bm{\tau}, \bm{\eta},\bm{\beta}, \phi\}$, where each of these blocks corresponding to a subproblem. Then, the two subproblems are solved iteratively to obtain the convergent solution.

Fig. \ref{AO_comp} compares the mEDT gap between the proposed IPCA algorithm and the AO algorithm versus the frequency of each helper, where the normalized gap is defined as the ratio of the mEDT gap to the mEDT achieved by the proposed IPCA algorithm. It can be observed that the normalized gap is approximately equal to $23.8\%$ and $15.3\%$ under $D_n = 2\times 10^5$ bites and $1\times 10^5$ bits, respectively, indicating that our proposed IPCA algorithm with alternating-free characteristics can achieve better performance. The reason is described as follows. Compared with the AO algorithm that decomposes PATACRA problem \eqref{eqCn} into two subproblems, our proposed IPCA algorithm solve it directly, thereby reducing some performance loss. On the other hand, the normalized gap is increasing with the data size. The reason is that when other parameters remain the same, the increase of the data size will lead to a smaller feasible domain, thus reducing the efficiency of both algorithms. Thus, it can be inferred that the proposed IPCA algorithm has better performance when the feasible domain is reduced.

\begin{figure}[!t]
\centering{\includegraphics[scale=0.5]{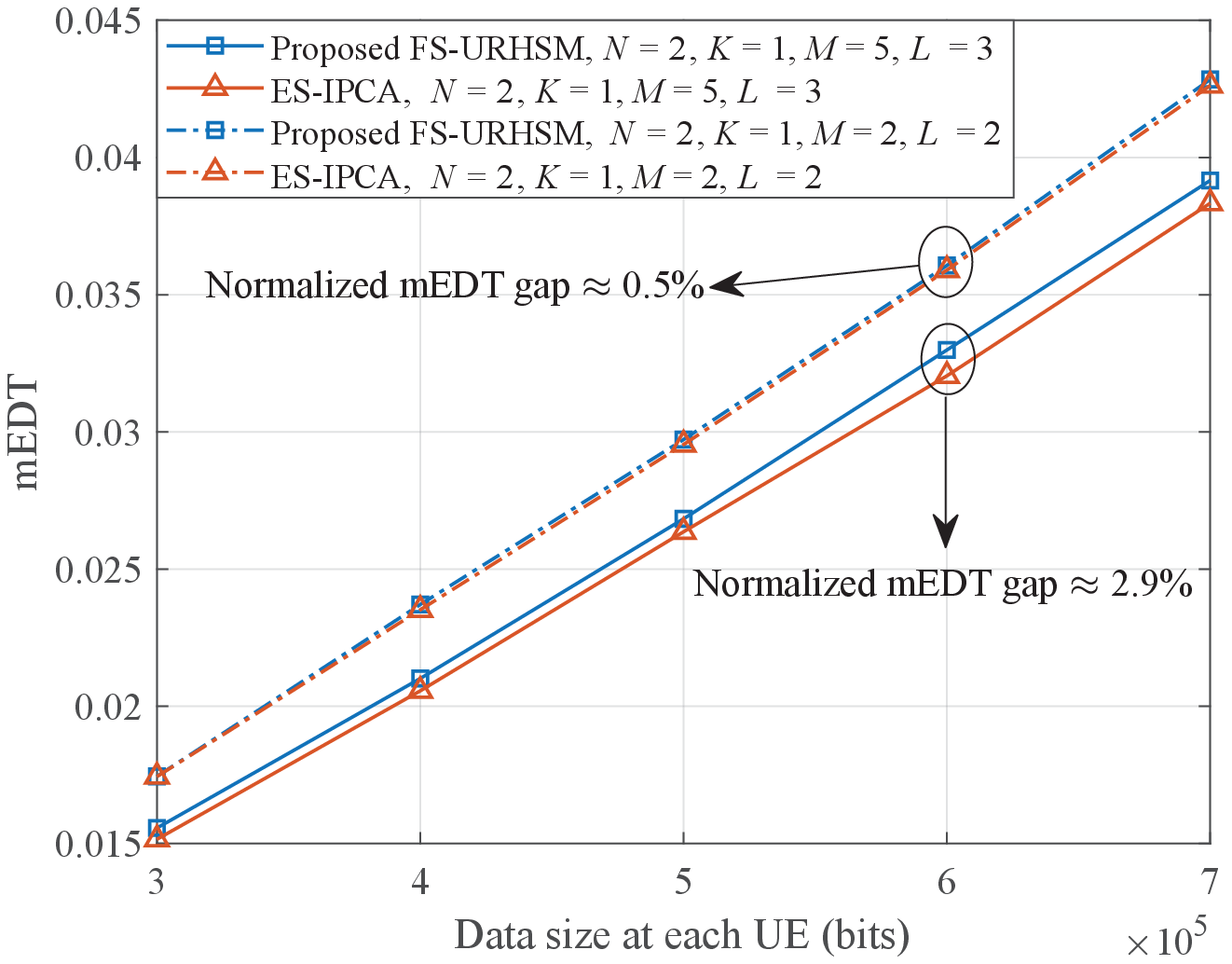}}
\caption{The mEDT comparisons of the proposed FS-URHSM algorithm and the ES-IPCA algorithm, where $\omega_{1}^E=0.6$ and $\omega_{2}^E = 0.3$. }
\label{EEfig1}
\vspace{-1em}
\end{figure}

To demonstrate the near-optimality of our proposed FS-URHSM algorithm for solving problem \eqref{UARA}, Fig. \ref{EEfig1} compares the performance gap between the proposed FS-URHSM algorithm and exhaustive search using the IPCA (called ES-IPCA) algorithm. The basic idea of the ES-IPCA algorithm is described as follows. By using the proposed IPCA algorithm to determine the mEDT of the possible matchings, the ES-IPCA algorithm searches all the possible matching conditions and selects the matching with the minimum mEDT.

Fig. \ref{EEfig1} compares the mEDT gap between the proposed FS-URHSM and the ES-IPCA algorithm versus the data size, where the normalized mEDT gap is defined as the ratio of the mEDT gap to the mEDT achieved by the ES-IPCA algorithm. As shown in Fig. \ref{EEfig1}, under $N=2$, $K=1$, $M=2$ and $L=2$, the normalized mEDT gap is approximately equal to $0.5\%$. When $N=2$, $K=1$, $M=5$ and $L=3$, the normalized mEDT gap is approximately equal to $2.9\%$, indicating that our proposed FS-URHSM algorithm achieves nearly the optimal solution. On the other hand, as shown in Fig. \ref{EEfig1}, the mEDT gap increases slightly when the number of MEC helpers and the RBs increase. This can be explained that the matching space between UEs, RBs, MEC helpers, and servers increases with the number of UEs/RBs/helpers/servers, thereby slightly degrading the performance of the proposed FS-URHSM algorithm. Recall that the worst-case complexity of the proposed FS-URHSM algorithm is $\mathcal{O}(\Delta\delta(4\delta_1 N^2+\delta_2 N(M-N)+\delta_3 N(L-N)+\delta_4 N(K-1)))$, which is much less than the exponential complexity of the ES-IPCA algorithm, i.e., $\mathcal{O}(\delta 2^{NM+NK+NL})$. Therefore, it can be concluded that the proposed FS-URHSM algorithm can achieve a near-optimal performance with much less complexity.
\subsection{Performance Comparisons }
Next, we compare our proposed FS-URHSM algorithm with the following schemes:
\begin{itemize}
\item \textit{FDMA-based offloading without Helpers (FDMA w.o. Helpers)}: Inspired by \cite{TCOM_CXC}, each UE associates with one server, divides its task into two subtasks, and offloads one subtask to the associated server while computing the remaining subtask locally.
\item \textit{NOMA-based offloading without helpers (NOMA w.o. Helpers)}: Inspired by \cite{FF_TCOM_imCSI}, each UE selects two MEC servers and divides its task into three subtasks. Then, each UE locally computes one of the subtasks while offloading the rest two subtasks to the selected servers by employing NOMA.
\item \textit{TDMA-based offloading with helpers (TDMA w. Helpers)}: Inspired by \cite{TDMAD2D6}, each UE selects one helper and one server, and divides its task into three subtasks. Then, each UE sequentially offloads two subtasks to the selected helper and the server while locally executing the rest subtask.
\item \textit{FDMA-based offloading with helpers (FDMA w. Helpers)}: Inspired by \cite{FDMAD2D1}, each UE selects one helper and one server, and divides its task into three subtasks. Then, each of them executes one of the subtasks locally and offloads the rest two subtasks to the selected one MEC helper and one server over two orthogonal RBs.
\end{itemize}
\begin{figure}[!t]
\centering{\includegraphics[scale=0.5]{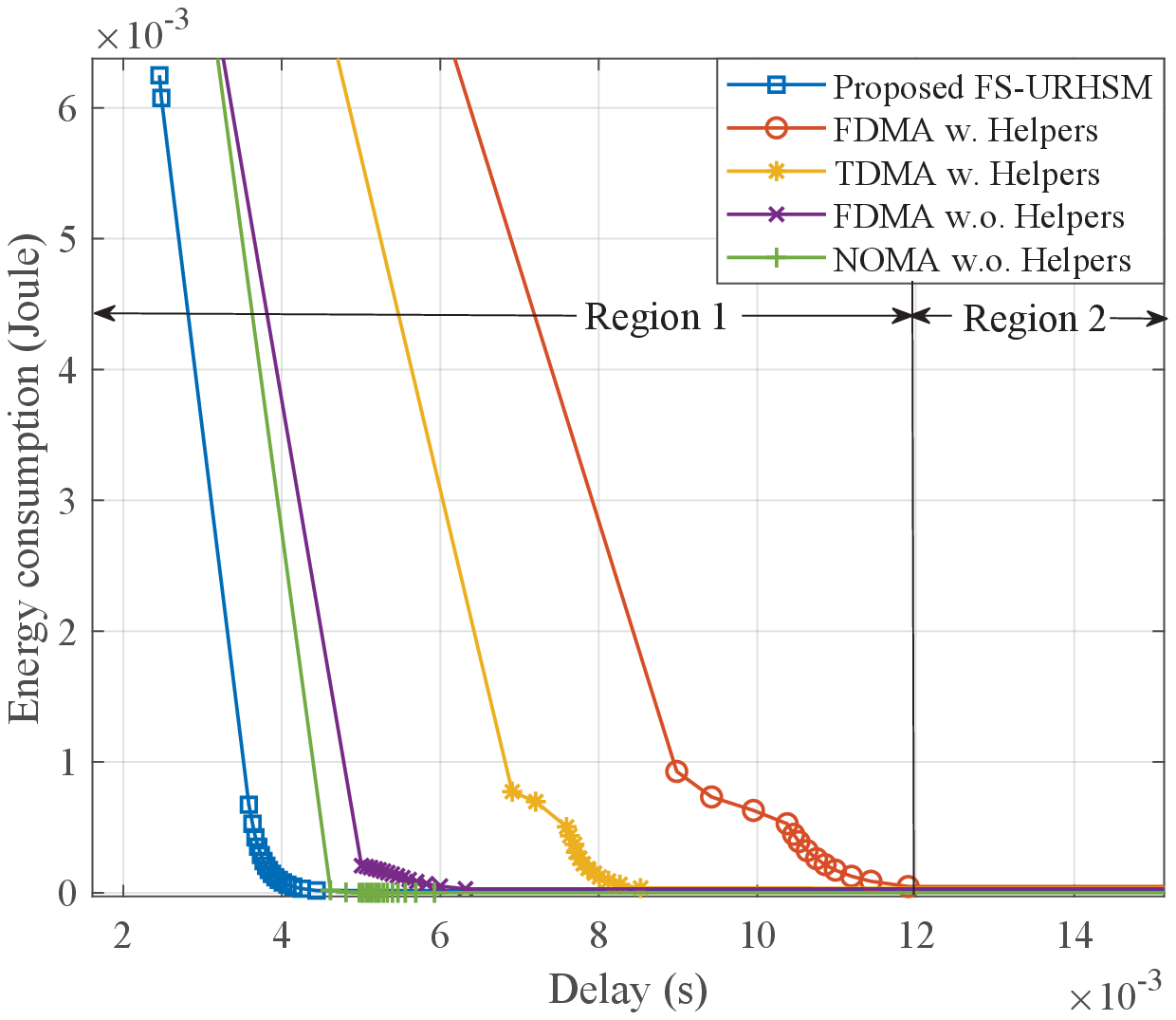}}
\caption{The tradeoff region among energy consumption and delay of the proposed FS-URHSM algorithm, FDMA w. Helpers, TDMA w. Helpers, FDMA w.o. Helpers and NOMA w.o. Helpers, where $\omega_{n}^E \in[0,1]$, $\omega_{n}^T=1-\omega_{n}^E$, $N=K=M=L =2$, and $D_n = 1\times 10^5$ bits. }
\label{verusw2}
\end{figure}

Fig. \ref{verusw2} describes the tradeoff between the energy consumption and the delay of proposed FS-URHSM algorithm and other schemes under different weight factors $\omega_{n}^E$ and $\omega_{n}^T$. As shown in Region 1 of Fig. \ref{verusw2}, the energy consumptions of the proposed FS-URHSM algorithm, FDMA w. Helpers scheme, TDMA w. Helpers scheme and FDMA w.o. Helpers scheme decrease with the delay, and our proposed FS-URHSM outperforms other schemes in terms of the delay when the energy consumption is small. This observation can be explained as follows. By using our FS-URHSM algorithm, the tasks can be simultaneously offloaded to the helper and the server, and performed in a parallel manner, indicating that our algorithm can reduce the energy consumption of the local computing and the delay of transmission more than other schemes. On the other hand, as shown in Region 2 of Fig. \ref{verusw2}, the proposed FS-URHSM algorithm consumes nearly the same energy of other schemes when the delay further increases to more than $0.012$ s. This is because when users become insensitive to the delay, each user is more likely to offload its entire task to the MEC server, in order to save the energy consumption incurred by local computing and NOMA enabled offloading. Based on the above observations, it can be concluded that our proposed scheme is more suitable for MEC systems with strict delay requirements.

\begin{figure}[!t]
\centering{\includegraphics[scale=0.5]{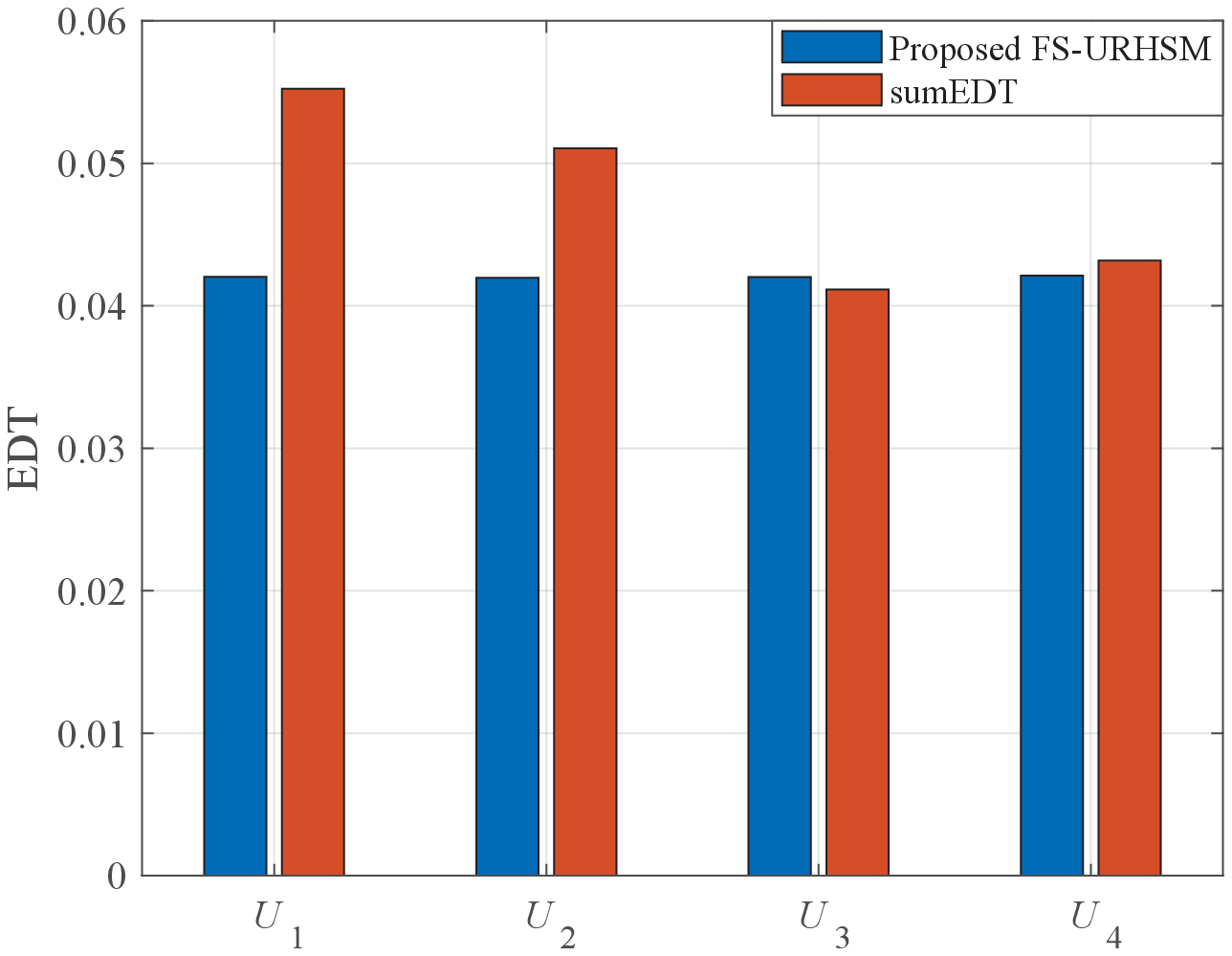}}
\caption{The fairness comparisons of the proposed FS-URHSM algorithm and the sumEDT, where $N=K=M=L=4$, $\omega_{n}^E = \omega_{n}^T = 0.5$, and $D_n = 1\times 10^6$ bits. }
\label{vssumEDT}
\end{figure}

Fig. \ref{vssumEDT} compares the user fairness between the proposed FS-URHSM algorithm and the sum of EDT (namely sumEDT) that minimizes the sum of all the UEs' EDT. It can be observed from Fig. \ref{vssumEDT} that the value of each UE's EDT obtained by our proposed FS-URHSM algorithm is nearly the same, while the value of each UE's EDT obtained by the sumEDT is significantly different. Moreover, the value of minimal EDT obtained by the sumEDT is lower than that obtained by our proposed FS-URHSM algorithm, indicating that our proposed FS-URHSM algorithm can achieve the min-max fairness among UEs at the cost of the EDT. This is because the goal of our proposed FS-URHSM algorithm is to minimize the maximal EDT among all UEs, while the sumEDT scheme aims to improve the performance of the UE that is more likely to reduce the energy consumption/delay.

\begin{figure}[!t]
\centering{\includegraphics[scale=0.5]{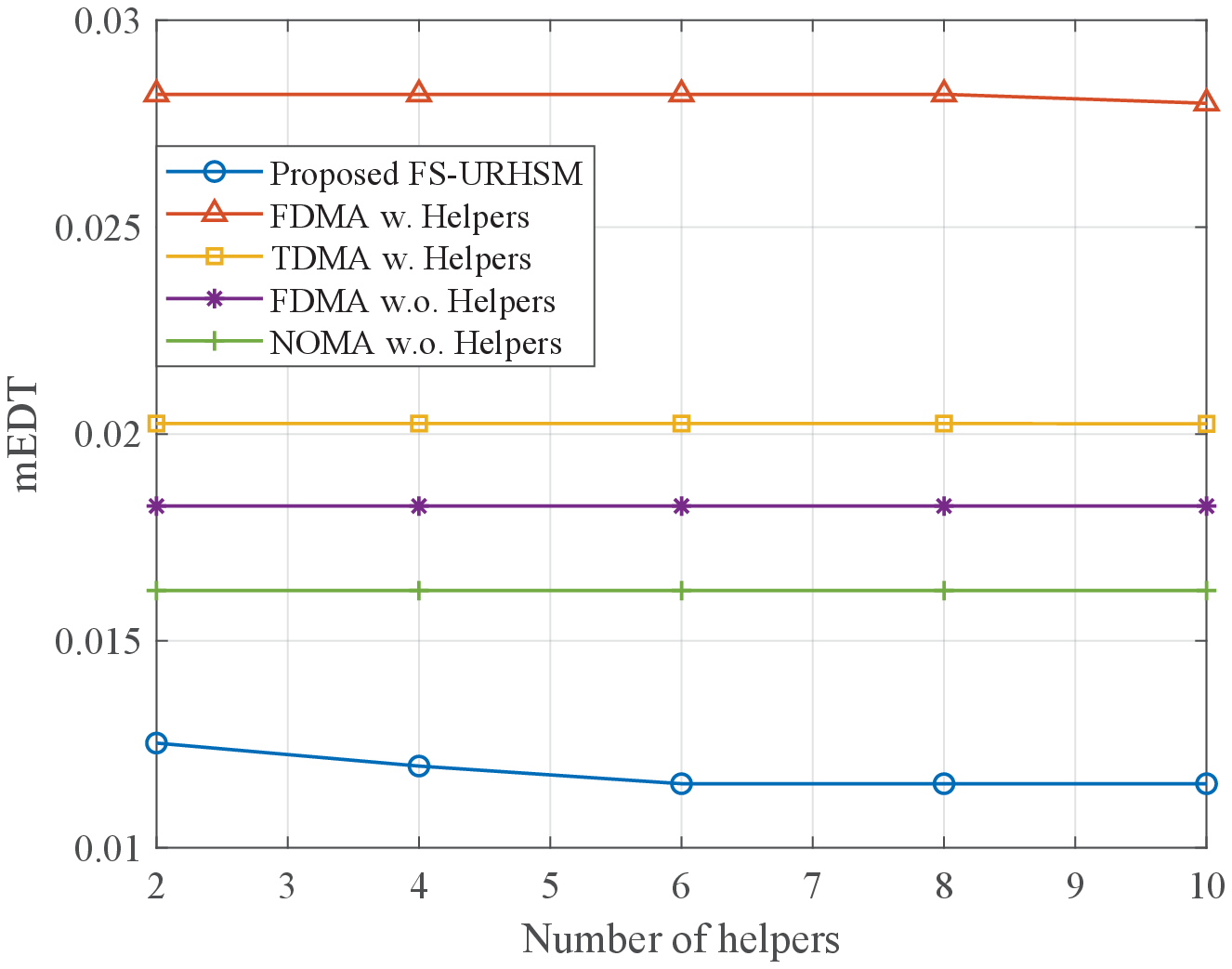}}
\caption{The mEDT comparisons of proposed FS-URHSM algorithm, FDMA w. Helpers, TDMA w. Helpers, FDMA w.o. Helpers and NOMA w.o. Helpers versus the number of helpers, where $N=K=L=2$, $M=10$, $D_n = 5\times 10^5$ bits, $\omega_{n}^E = 0.5$, and $\omega_n^T = 1-\omega_{n}^E$.}
\label{vsHelpers}
\end{figure}

\begin{figure}[!t]
\centering{\includegraphics[scale=0.5]{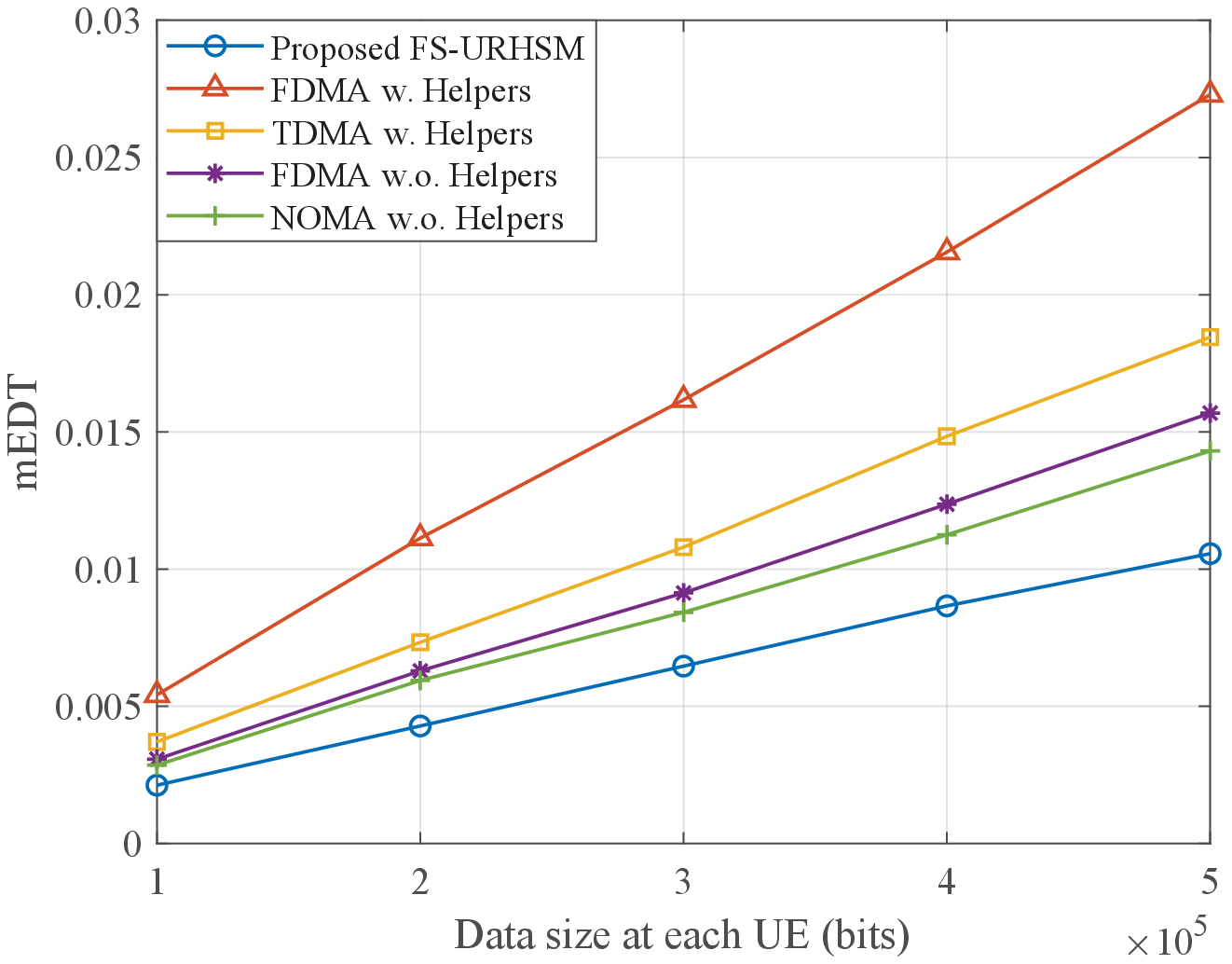}}
\caption{The mEDT comparisons of proposed FS-URHSM algorithm, FDMA w. Helpers, TDMA w. Helpers, FDMA w.o. Helpers and NOMA w.o. Helpers schemes versus data size at each UE,  where $N=K=M=L=2$, $\omega_{n}^E = 0.5$, and $\omega_n^T = 1-\omega_{n}^E$.}
\label{vsData}
\vspace{0em}
\end{figure}

Figs. \ref{vsHelpers} and \ref{vsData} depict the mEDT of the proposed FS-URHSM algorithm and other schemes versus the data size and the number of MEC helpers, respectively. It can be observed from all the figures that the proposed FS-URHSM algorithm outperforms other schemes in terms of the mEDT.
This observation can be explained as follows.
1) By properly choosing helpers, MEC servers and RBs, the computation loads of UEs can be transferred to the helpers and the servers that have better computation capabilities than users. Therefore, the energy consumption of local computing and the delay of task completion can be reduced.
2) Although UEs should consume more energy to handle the interference incurred by downlink NOMA-enabled offloading, they can offload/compute tasks in a parallel manner, thus largely reducing the transmission/computation delay.
3) Since the UEs can harvest the spare resources of helpers by using our proposed algorithm, the transmission delay and energy consumption can be further reduced compared with the NOMA w.o. Helpers scheme, and the computation workload on the servers can also be reduced.
\section{Conclusion}
\label{Section_conclusion}
This paper has investigated a communication and computation resources allocation scheme in a helper-assisted NOMA-MEC system.
We have formulated the joint optimization of user association, RB assignment, power allocation, task assignment, and computation resource allocation as a mixed-integer nonlinear programming problem, in order to minimize the mEDT. To efficiently solve this problem, the formulated problem has been decoupled as a PATACRA subproblem, and then the original problem has been equivalently reformulated as a DUARA problem with the solution obtained from the PATACRA subproblem. We have proposed an IPCA algorithm to solve the PATACRA subproblem, with which a low-complexity FS-URHSM algorithm has been proposed to solve the DUARA problem. Numerical results have demonstrated that the proposed FS-URHSM algorithm achieves the near-optimal solution with much less complexity, achieves the min-max fairness among UEs, and performs better than other schemes in terms of mEDT.

\appendices
\renewcommand{\theequation}{\thesection.\arabic{equation}}
\section{Convexity Proof of Constraint \eqref{cons_CCRA_re2_EDT}}
\setcounter{equation}{0}
Function $E_n^\prime$ has terms $\tau_{n}^{\text{tran}} 2^{(\eta_{n}^S{D_n})/(B \tau_{n}^\text{tran})}$ and $\tau_{n}^{\text{tran}}   {2^{ ({\eta_{n}^H{D_n} + \eta_{n}^S{D_n}})/({B \tau_{n}^{\text{tran}}})}} $ with the perspective structure given by
\begin{align}
\label{eqpers}
f = y2^{x/y},
\end{align}
where $x,y>0$. Hence, the second-order Hessian matrix of \eqref{eqpers} can be expressed by
\begin{align}
\label{H}
\mathcal{H} = \frac{1}{(\ln 2)^2 y}
\begin{bmatrix}
2^{\frac{x}{y}}   &    -{\frac{x}{y}2^{\frac{x}{y}}} \\
 -{\frac{x}{y}2^{\frac{x}{y}}}  & ({\frac{x}{y})^2 2^{\frac{x}{y}}}
\end{bmatrix}.
\end{align}
For all $\bm{v}=[v_1, v_2]^H \in \mathbb{R}^2$, Hessian matrix $\mathcal{H}$ satisfies
\begin{align}
\bm{v}^H \mathcal{H} \bm{v} = \frac{1}{(\ln 2)^2 y} (v_1-v_2 {\frac{x}{y}})^2 2^{\frac{x}{y}} \ge 0.
\end{align}
Therefore, $\mathcal{H}$ is a positive semi-definite matrix and \eqref{eqpers} is joint-convex with respect to variables $x,y$. Furthermore, recall that  $A_{n,1} \ge 0, A_{n,2} \ge 0$ and the rest terms in the left hand of constraint \eqref{cons_CCRA_re2_EDT} are affine functions. The left hand of constraint \eqref{cons_CCRA_re2_EDT} is a linear combination of convex and affine functions, while the right hand of constraint \eqref{cons_CCRA_re2_EDT} is an affine function. As a result, \eqref{cons_CCRA_re2_EDT} is a convex constraint.
\ifCLASSOPTIONcaptionsoff
  \newpage
\fi

%





\end{document}